\documentclass[a4paper,12pt]{article}

\usepackage{pstricks}
\usepackage{graphicx,psfrag}

\usepackage[centertags]{amsmath}
\usepackage{amsfonts}
\usepackage{amssymb}
\usepackage{amsthm}
\usepackage{newlfont}
\usepackage{graphicx}
\usepackage[active]{srcltx}

\usepackage{enumerate}

\newtheorem{theorem}{Theorem}

\newtheorem{lemma}{Lemma}

\theoremstyle{remark}

\theoremstyle{definition}

\theoremstyle{definition}

               \newcommand {\Lam}  {\Lambda}

      \newcommand {\I}  {{R}}
      \newcommand {\LL}  {{S}}

\title{High Frequency Scattering by a Classically Invisible Body}

\author{ E.Lakshtanov\thanks{Department of Mathematics, Aveiro University, Aveiro 3810, Portugal.  This work
was supported by {\it Centre for Research on Optimization and
Control} (CEOC) from the ''{\it Funda\c{c}\~{a}o para a
Ci\^{e}ncia e a Tecnologia}'' (FCT), cofinanced by the European
Community Fund FEDER/POCTI, and by the FCT research project
PTDC/MAT/113470/2009 (lakshtanov@rambler.ru).} \and
B.D.Sleeman\thanks{School of Mathematics, University of Leeds, Leeds, LS2 9JT,UK (bds@maths.leeds.ac.uk)} \and
 B.Vainberg\thanks{Department
of Mathematics and Statistics, University of North Carolina,
Charlotte, NC 28223, USA. The work was partially supported  by the NSF grant DMS-1008132 (brvainbe@uncc.edu).}}

\begin{document}
\maketitle

\begin{abstract}
We consider a polyhedron with zero classical resistance, i.e., a
polyhedron invisible to an observer viewing only the paths of
geometrical optics rays. The corresponding problem of
scattering of plane waves by the polyhedron is studied. The
quasiclassical approximation is obtained and justified in the case
of impedance boundary conditions with a non zero absorbing part.
It is shown that the total momentum transmitted to the obstacle
vanishes when the frequency $k$ goes to infinity, and that the total
cross section oscillates at high frequencies. When the impedance
$\lambda_0$ is real (i. e., there is no absorption), it is shown that there
exists a sequence of frequencies $k_n$ such that
the averages in the impedance of the total cross section over shrinking intervals around $\lambda_0
$ go to zero as $k_n \to  \infty$.
\end{abstract}



\section{Introduction}

An interesting geometrical object was studied in a recent publication by Aleksenko and Plakhov \cite{alex1}. This object
 $\mathcal O$ has the following property. Geometrical optical rays, coming from a particular direction and reflected
twice from the boundary of $\mathcal O$ by the law of geometrical optics, continue to propagate parallel to each other in the
same way as if the obstacle was absent. The object has zero classical total scattering cross section and appears
invisible to an observer on the basis of the theory of geometrical optical rays. Note that a phase shift
 may influence the "invisibility" of the obstacle. One should also note that optical ray considerations provide an approximation to
the expected properties of the corresponding optical problem, when the obstacle is smooth and convex. These conditions do
not hold for the object under consideration. A rigorous treatment of the problem has to be based on an
investigation of the solutions of the wave equation.

This paper concerns the study of the associated scattering problem for the
reduced wave (i. e., Helmholtz) equation. High frequency
asymptotics are obtained for the scattering of plane wave by the Aleksenko-Plakhov
obstacle $\mathcal O$. The rigorous
justification will be obtained in the case of impedance
boundary conditions with a non-zero absorption. It will also be
obtained in a weaker sense for boundary conditions without
 absorption. It is
important to note that the delay time is the same for all the rays meeting
the obstacle, i.e., the phase shift $\Delta$ is a constant for all
rays meeting $\mathcal O$. In the absence of absorption, this implies that the
obstacle $\mathcal O$ is almost invisible at a sequence of high frequencies $k=k_n=\frac{k^0+2\pi n}{\Delta},~n\to\infty$,
where $k^0$ is determined by the boundary condition, and the invisibility effect disappears for other frequencies.

Note that the scattering theory prohibits the existence of
absolutely invisible bodies, since a nontrivial outgoing solution of the Helmholtz equation can not have zero scattering amplitude.
 Our results show that the scattering data for
one incident direction and a sequence of frequencies can be as
small as we please for an obstacle of an arbitrary size. Another
by-product of our results concerns the relation between the
total scattering cross section and the geometrical cross section.
For the obstacle under consideration, the total scattering cross
section approaches to four times the geometrical cross section for some increasing sequence of frequencies (when the incident and the reflected waves are
in-phase), while for convex obstacles the corresponding ratio
approaches to two as the frequency increases.

The simplicity of the model under consideration allows one to consider
it as a candidate in competition with other models on cloaking
(eg. \cite{Norris}). Its obvious disadvantage, namely the incident
direction is fixed, may be non-essential in some applications. On
the other hand, the corresponding cloaking object is easy to make, in contrast to other
 cloaking materials which present formidable
technological difficulties. Indeed currently proposed cloaking materials which lead to
zero total scattering cross section for all incident directions need
to have an infinite mass \cite{Norris} or to be highly
non-isotropic \cite{1, ss}. The latter leads to huge difficulties
in a practical realization of such a non-isotropic coating which
are not attainable at present.

\begin{figure}[h]
\begin{picture}(0,110)
 \scalebox{0.5}{ \rput(11.6,3.5){
\pspolygon[fillstyle=solid,fillcolor=gray,linewidth=1.6pt](-4,-3.4641)(-3.7,0)(-4,3.4641)(-2,0)
   \pspolygon[fillstyle=solid,fillcolor=lightgray,linewidth=1.6pt](-4,3.4641)(-2,0)(-1.1,0.3)(-2.8,3.2641)
   \pspolygon[fillstyle=solid,fillcolor=lightgray,linewidth=1.6pt](-4,-3.4641)(-2,0)(-1.1,0.3)(-2.8,-2.6641)
\pspolygon[fillstyle=solid,fillcolor=gray,linewidth=1.6pt](4,-3.4641)(3.7,0)(4,3.4641)(2,0)
   \pspolygon[fillstyle=solid,fillcolor=lightgray,linewidth=1.6pt](4,-3.4641)(3.7,0)(4.7,0.3)(4.9,-2.6641)
   \pspolygon[fillstyle=solid,fillcolor=lightgray,linewidth=1.6pt](3.7,0)(4,3.4641)(4.9,3.2641)(4.7,0.3)

   \rput(5.2,-2.8){\small D}
   \rput(4.1,-3.8){\small B}
\rput(-4.1,-3.8){\small A}

} }
\end{picture}
 \caption{Aleksenko-Plakhov object $\mathcal O$. Translation
distance $|BD|=1$. }
\label{fig1}
\end{figure}
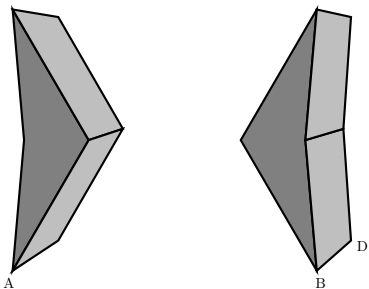

The Aleksenko-Plakhov obstacle $\mathcal O \subset \mathbb R^3$ with Lipschitz
boundary $\partial \mathcal O$ considered in this paper is
presented in Fig.~\ref{fig1}. It is formed by an
orthogonal translation of the 2D object shown in Fig.~\ref{fig2}. The
prolongation of this translation is $1$. In Fig.~\ref{fig2} $A'C'B'$ and $ACB$ are equilateral triangles.
where $C$,$C'$ are the mid points of segments $A'B'$ and $AB$,
respectively. The lateral sides $A'A$ and $B'B$ are slightly displaced in
order to avoid a positive measure of the obstacle's surface on the
tangent rays propagating along the $z-$axis. We suppose that
$A'B'=1$ that means that geometrical cross section of the body
i. e.,the area of the projection of the body on the $(x,y)$-plane,
equals to $1/2$.

The geometry of this obstacle is chosen in such a way that the
geometrical optical rays propagating from the $z$-direction, after a double
reflection from the boundary of the obstacle, continue to move
parallel to each other with a constant shift $\Delta$ of the phase. This
creates the effect of invisibility for an observer who relies only on geometrical optical
rays coming from above (we discuss this in more detail in the next section). In fact, the
geometry of triangles $A'C'B'$ and $ACB$ can be chosen more generally, and other
 obstacles also can be constructed which also create
the same effect of classical invisibility \cite{alex1}.

The goal of this paper is to study this invisibility effect using a
rigorous analysis based on the investigation of the corresponding
wave problem. It is well known that the justification of the
geometrical optics approximation is a difficult task. This was done
earlier in two cases: for smooth strictly convex obstacles
\cite{majda, mel, ML, Ludwig, babich} and for inhomogeneous media
in the whole space \cite {vai}. For example in the first case, a
complicated approximation of the field near the tangent rays is
needed, and delicate local energy estimates are used. The latter
are known for strictly convex or star-shaped obstacles. Some
conditional results  for non-convex obstacle can be found in
\cite{pet} where the scattering amplitude is studied along the
non-caustic directions.

Let $k > 0$, the scattered field $u$
is a solution of the Helmholtz equation satisfying the Sommerfeld radiation condition:
\begin{align}\label{helm}
\Delta u(r)+k^2 u(r)=0, \quad r \in \Omega=\mathbb R^3 \backslash
\mathcal O,
\end{align}
\begin{align}\label{Somm}
\int_{|r|=R} \left |\frac{\partial u(r)}{\partial |r|}-iku(r)
\right |^2 dS = o(1), \quad R \rightarrow \infty.
\end{align}
We assume an impedance boundary condition holds on $\partial
\mathcal O$, i.e.
\begin{align}\label{bcOmega}
\left (\frac{\partial}{\partial n} + k \lambda \right )(u+e^{ik(r
\cdot p_0)}) = 0, \quad \quad r=(x,y,z) \in
\partial \mathcal O, \quad p_0=(0,0,1),
\end{align}
where $\Im \lambda \geq 0$ is a constant and $n$ denotes the outer unit
normal for $\mathcal O$. There exists a unique solution in
$H^1_{loc}(\Omega)$ which satisfies all these conditions (eg
\cite{Mcl}). Every solution $u(r)$  of
(\ref{helm})-(\ref{bcOmega})   has the following behavior at
infinity
\begin{equation}\label{scamp}
u(r)=\frac{e^{ik|r|}}{|r|} u_{\infty}(\theta)+o \left
(\frac{1}{|r|} \right ), \quad r \rightarrow \infty, \quad
\theta=r/|r| \in S^2,
\end{equation}
where the function $u_\infty(\theta)=u_\infty(\theta,k)$ is called
the {\it scattering amplitude} and the quantity
\begin{equation}\label{tcrs}
\sigma(k)=\|u_\infty\|^2_{L_2(S^2)}=\int_{S^2}
|u_\infty(\theta)|^2 d\mu (\theta)
\end{equation}
is called the {\it total cross section}. Here $d\mu$ is the surface element of
the unit sphere.

The following observable is called the {\it transport cross section}.
$$
\sigma_T(k)= \int_{S^2} (1-\theta \cdot p_0))|u_\infty(\theta)|^2
d\mu (\theta)
$$
and equals the total momentum transmitted
to the obstacle per unit time in large volume normalization.

The following are the main results obtained in this paper. They are stronger if $\Im \lambda>0$ (i. e., there is absorption of the energy at the boundary),
and hold on average if $\Im \lambda=0$.
\begin{theorem}\label{tint}
Let $\Im \lambda>0$. Then

1) the transport cross section vanishes as $k$ goes to infinity:
$$
\lim_{k \rightarrow \infty} \sigma_T=0,
$$

2) the total cross section  has the following asymptotic behavior for large $k$:
\begin{equation}\label{st2}
\sigma(k)= \frac{1}{2} \left |A^2e^{ik\Delta}-1 \right |^2 +o(1),~~
\quad k \rightarrow \infty,~~A=\frac{i-2\lambda}{i+2\lambda},
\end{equation}
where $\Delta$ is defined in Fig.\ref{fig2}.
\\
3) $|u_\infty (\theta)|^2 \sim \sigma(k) \delta(p_0),
\quad k \rightarrow \infty$
in the sense of distributions, i.e.,  for any $\varphi \in C(S^2)$, we have
\begin{equation}\label{DistrLimitO}
\int_{S^2} \varphi(\theta) |u_\infty (\theta)|^2 dS_\theta =
\frac{1}{2} \left |-1+A^2e^{ik\Delta} \right |^2 \varphi(p_0) +
o(1), \quad k \rightarrow \infty.
\end{equation}
\end{theorem}
{\bf Remark.} If $\lambda$ is real, then $|A|=1$. Thus if the second statement holds for real $\lambda$ then
$$
\sigma(k)= \frac{1}{2} \left |-1+e^{i(k\Delta+2 \arg (A))}
\right |^2 +o(1), \quad k \rightarrow \infty.
$$
and
$$
\lim_{n \rightarrow \infty} \sigma(k_n)=0~~\text{for}~ k_n=\frac{-2\arg(A)}{\Delta}+\frac{2\pi}{\Delta}n,~~n \in
\mathbb Z,~~\Im \lambda =0.
$$
The almost invisibility of $\mathcal O$ manifests itself by the fact that the total cross section $\sigma(k)$ can be made
as small as we please by choosing a complex impedance $\lambda$ close enough to an arbitrary real $\lambda=\lambda_0$ and
then choosing $k=k_n$ large enough.

It also follows from (\ref{st2}) with real $\lambda$ that $\sigma(k)\rightarrow 2$ when
$k=\frac{-2\arg(A)}{\Delta}+\frac{(2n+1)\pi}{\Delta} \rightarrow \infty$, i.e., $\sigma(k)$ approaches four times the
 geometrical cross section. The latter is the area of the shadow of the obstacle if the reflected rays are not
taken into account, and this area equals $1/2$. For convex obstacles, the ratio of the total to the geometrical cross
 section approaches two as $k \rightarrow\infty$.

In some cases we will denote the total cross section (\ref{tcrs}) by $\sigma_{\lambda}(k)$ in order to stress its dependence on the value of the parameter $\lambda$ in the boundary condition (\ref{bcOmega}).
\begin{theorem}\label{tint1}
Let $\lambda_0$ be real and
\begin{equation}\label{kfrom}
k_n=\frac{-2\arg(A(\lambda_0))}{\Delta}+\frac{2\pi}{\Delta}n, \quad n
\in \mathbb Z.
\end{equation}
Then there exists a sequence of positive numbers $\varepsilon_n\rightarrow 0,~n\rightarrow\infty,$ such that
\[
\frac{1}{\varepsilon_n}\int_{\lambda_0-\varepsilon_n}^{\lambda_0+\varepsilon_n}\sigma_{\lambda}(k_n)d\lambda\rightarrow 0,~~n\rightarrow\infty.
\]
\end{theorem}
\textbf{Consequence.} Under conditions of Theorem \ref{tint1}, there exists a sequence of real
numbers $\lambda_n \rightarrow \lambda_0$ such that $\lim_{n \rightarrow \infty} \sigma_{\lambda_n}(k_n)=0.$

The paper is organized as follows. The eikonal geometrical optics
approximation is constructed in the next section. It provides a
basis to understand the validity of Theorems \ref{tint},
\ref{tint1}. It is also needed in order to construct the Kirchhoff
approximation $u^0$ which is used to prove these results. Section
3 outlines the proof of Theorem \ref{tint}. There we state the
asymptotic properties of the Kirchhoff
 approximation in a bounded region and discuss the estimates needed to justify the approximation. In order to obtain
the properties of $u^0$ we study first Kirchhoff approximation for the problem of scattering by a single polygon
(one face of the obstacle $\mathcal O$). This is done in section 4 with the main technical parts of the proof moved
into Appendix 1. The asymptotic behavior in a bounded region of the Kirchhoff  approximation for the obstacle $\mathcal O$
is obtained in section 5. Sections 6 and 7 provide the asymptotic behavior of the total and transport cross sections,
 respectively, when $\Im \lambda>0$. The case $\Im \lambda=0$ is studied in section 8.

\section{Eikonal approximation}\label{eik}
We begin by recalling that the construction of the eikonal approximation to the
solution of the  scattering problem uses a Lagrangian manifold
$\Lambda$ formed by trajectories $(r,p)=(r(t),p(t))$ in the phase
space $\mathbb R^3 \times \mathbb R^3$ which correspond to the
geometrical optics rays $r=r(t)$. The components of the vector
$(r,p)$ are the position $r=r(t)$ of a point along the ray at time $t$ and the momentum $p=p(t)$ (where $p$ is the unit
 vector along the ray). The trajectories start at time $t=0$ at points $(x_0,y_0,-a)$ of the plane $z=-a$ with
momentum $p_0=(0,0,1)$. It is supposed that the obstacle $\mathcal O$ is
situated in the half space $z>-a$. Location and momentum $(r,p)$ of
every ray, except those which meet an edge of $\mathcal O$, are determined
uniquely for all $t \in \mathbb R$. Thus, $(x_0,y_0,t) \in \mathbb
R^3$
 are global coordinates on $\Lambda$. Let
  $P_r : \Lam \rightarrow \mathbb R^3$ be the  projection of $\Lambda$ to the
 coordinate space $\mathbb R^3$ and let $\mathcal B =P_r \partial \Lam \subset \mathbb R^3$.

In the case of the Neumann boundary condition, the eikonal geometrical optics
approximation has the form:
\begin{equation}\label{app1}
\Psi^{eik}(r)=\sum_{(x_0,y_0,t): r(x_0,y_0,t)=r}\left |
\frac{D(r)}{D(x_0,y_0,t)}\right |^{-1/2}e^{ikS(x_0,y_0,t)}, \quad
r \in \mathbb R^3 \backslash \mathcal B.
\end{equation}
Here the action $S$ is just the length of
the trajectory, i.e. $S(x,y,t)=t$ (since the exterior of
the obstacle is homogeneous), the Jacobian $|D(r)/D(x_0,y_0,t)|$ is the density of geometrical optics rays in ray tubes,
 and the summation is taken over all the points of $\Lambda$
having $r$ as projection to $\mathbb R^3$.

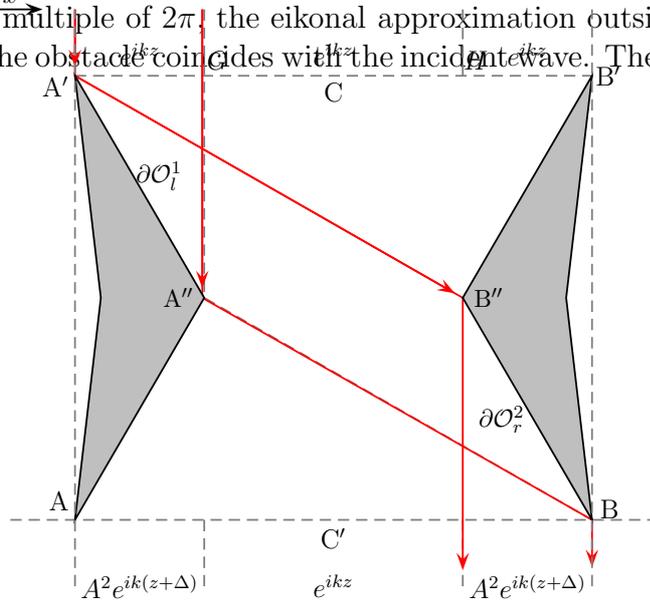
\begin{figure}[h]
\begin{picture}(0,175)
\scalebox{0.85}{ \rput(8.7,4.3){

\psline[linecolor=gray,linestyle=dashed](-2,4.5)(-2,3.4641)(-2,0)(4,-3.4641)

\psline[linecolor=gray,linestyle=dashed](2,4.5)(2,3.4641)
\psline[linecolor=gray,linestyle=dashed](2,-3.4641)(2,-4.5)
\psline[linecolor=gray,linestyle=dashed](-2,-3.4641)(-2,-4.5)

\rput(3,3.8){$e^{ikz}$}
\rput(0,3.8){$e^{ikz}$}
\rput(-3,3.8){$e^{ikz}$}

\rput(3,-4.5){$A^2e^{ik(z+\Delta)}$} \rput(0,-4.5){$e^{ikz}$}
\rput(-3,-4.5){$A^2e^{ik(z+\Delta)}$}

\psline[linecolor=gray,linestyle=dashed](-5,-3.4641)(5,-3.4641)
\psline[linecolor=gray,linestyle=dashed](-4,3.4641)(4,3.4641)

 \psline[linecolor=red,arrows=->,arrowscale=1.8](-4,4.5)(-4,3.6)
\psline[linecolor=red,arrows=->,arrowscale=1.8](-4,4.5)(-4,3.4641)(1.88,0.069282)
\psline[linecolor=red,arrows=->,arrowscale=1.8](-4,3.4641)(2,0)(2,-4.25)
 \psline[linecolor=red,arrows=->,arrowscale=1.8](-2.03,4.5)(-2.03,0.15)
\psline[linecolor=red,arrows=->,arrowscale=1.8](-2.03,4.5)(-2.03,0)(4,-3.4641)(4,-4.2)

\pspolygon[fillstyle=solid,fillcolor=lightgray](-4,-3.4641)(-3.6,0)(-4,3.4641)(-2,0)
\pspolygon[fillstyle=solid,fillcolor=lightgray](4,-3.4641)(3.6,0)(4,3.4641)(2,0)
\psline[linecolor=gray,linestyle=dashed](-4,-4.5)(-4,4.5)
\psline[linecolor=gray,linestyle=dashed](4,-4.5)(4,4.5)

\rput(-4.25,-3.1641){A} \rput(-4.3,3.3){A$'$}
  \rput(-2.7,1.9){\small $\partial \mathcal O^1_l$}
\rput(4.27,-3.3){B}
\rput(4.25,3.464){B$'$}
\rput(-1.8,3.7){$G$}
\rput(2.2,3.7){$H$}
\rput(-2.4,0){\small A$''$}
\rput(2.4,0){\small B$''$}
   \rput(2.6,-1.9){\small $\partial \mathcal O^2_r$}
   \rput(0,3.2){C}
   \rput(0,-3.75){C$'$}

 \psline[arrows=->,arrowscale=1.8](-5.5,4.5)(-4.5,4.5)
 \rput(-5,4.7){$x$}
 \psline[arrows=->,arrowscale=1.8](-5.5,4.5)(-5.5,3.46)
 \rput(-5.35,4){$z$}
} }
\end{picture}
 \caption{Plane section of
$\mathcal O$. Base side $|A'B'|=1$. Eikonal approximation $\Psi^{eik}$ is given above $A'B'$ and below $AB$; $~\Delta=|GA''|+|A''B|-|A'A|$.}
\label{fig2}
\end{figure}

Function $\Psi^{eik}$ satisfies
(\ref{bcOmega}) outside $\mathcal B$ and is discontinuous  on
$\mathcal B$. Nevertheless, we can use it to predict quasi-classic
effects. Note that reflection from the plane faces of $\mathcal O$ does not change the
density of the rays and so
\begin{align}\label{app1B}
\Psi^{eik}(r)=\sum_{(x_0,y_0,t): r(x_0,y_0,t)=r} e^{ikt}, \quad r
\in \mathbb R^3 \backslash \mathcal B.
\end{align}
In Figure~\ref{fig2} (where $A=1$ in the case of Neumann or $A=-1$ in the
case of Dirichlet boundary conditions) we see that the obstacle
does not have a shadow: for example, the shadow zone of the right
part of the obstacle is covered by the rays reflected from $A'A''$
and $B''B$.

The following observation on the geometrical structure is
important: each trajectory started on the line $A'B'$ between
points $A'$ and $G$ or between point $H$ and $B'$ and ended on the
line $AB$ have the same length.
It means that   all the rays below the line $AB$ have momentum
$p_0=(0,0,1),$ the action for the rays coming through $GH$ is $z$,
and the action for the rays which have collisions with  the
obstacle is $\Delta+z$, where $\Delta=|GA''|+|A''B|-|A'A|$.  So
the rays below  the line $AB$ differ from the sets of rays in the absence of the obstacle only by
the phase shift $\Delta$ on the rays coming through $A'H\cup GH'$. Hence, if $k\Delta$ is a multiple of
$2\pi$, the eikonal approximation outside of some neighborhood of the obstacle coincides with the incident wave.
The scattering amplitude is not defined for the eikonal
approximation, however the transport cross section for the eikonal
approximation is defined by the following limit:
$\lim_{R\rightarrow\infty}\int_{|r|=R} (1-\theta\cdot p_0)
|\Psi^{eik}(r)-e^{ikr\cdot p_0}|^2dS.$

Since all rays have forward direction after
collisions with the obstacle, the transport cross section
for the eikonal approximation is zero.
The arguments above concern the case of the Neumann boundary
condition ($\lambda=0$).  When the impedance $\lambda$ is
arbitrary, the terms in the eikonal approximation (\ref{app1}) after each
collision are multiplied  by the  factor
\begin{equation}\label{factor}
\frac{in_{\alpha}+\lambda}{in_{\alpha}-\lambda},
\end{equation}
where $n_{\alpha}=n\cdot\alpha$ is the cosine of the angle between
the normal to the surface at the point of the incidence and  the
direction $\alpha$ of the ray before the incidence. In the case of
our obstacle $ \mathcal O$ all incidences occur with the
same $n_{\alpha}=-1/2.$

Define
$
A(\lambda):=
\frac{i-2\lambda}{i+2\lambda},
$
then the eikonal approximation in the case of the impedance
boundary condition can be  written as
\begin{equation}\label{app1im}
\Psi^{eik}(r)=\sum_{(x_0,y_0,t): r(x_0,y_0,t)=r}
[A(\lambda)]^{n(x_0,y_0,t)}e^{ikt}, \quad r \in \mathbb R^3 \backslash
\mathcal B.
\end{equation}
where $n(x_0,y_0,t)$ denotes the number of collisions of the ray
with initial coordinates $(x_0,y_0)$ with $\partial \mathcal O$
before time $t$.
Thus the field  behind the triangles (after two collisions) is
$A^2e^{ik(z+\Delta)}$ (see Fig.~\ref{fig2}).

Note that the factor (\ref{factor}) is chosen by the requirement for $\Psi^{eik}(r)$ to satisfy the impedance boundary condition, i.e.
\begin{equation}\label{bcOmega11}
\left (\frac{\partial}{\partial n} + k \lambda \right )\Psi^{eik}(r) = 0, \quad \quad r \in
\partial \mathcal O.
\end{equation}

\section{Outline of the proof of Theorem \ref{tint}}
 The main technical
difficulties of the present paper concern the construction of an
appropriate asymptotic approximation for the scattered field. We
will not work directly with $\Psi^{eik}$ which has jumps and does not satisfy equation (\ref{helm}). Instead we  use the
 Kirchhoff
approximation $u^0$ to the solution $u$ in order to prove Theorem \ref{tint1}. This necessitates the need to justify the validity of
 the Kirchhoff approximation and to study
its asymptotic behavior.

Recall that the Green formula allows one to represent field  $u$ in terms of a surface integral involving
$u|_{\partial \mathcal O}, \frac{\partial u}{\partial
n}|_{\partial \mathcal O}$. If $\partial \mathcal O$ were smooth, the Kirchhoff approximation would be given by the same
Green formula with $u$ and its derivative in
the integrands are replaced by $\psi^{eik}(r):=\Psi^{eik}(r)-e^{ikr\cdot p_0}$ and
its derivative, respectively, i.e.,
\begin{equation*}
u^0\approx\frac{1}{4\pi}\int_{\partial \mathcal O}
[-\frac{\partial}{\partial n_q} ( \psi^{eik}(q)) \frac{e^{ik|r-q|}}{|r-q|} +
\psi^{eik}(q)\frac{\partial}{\partial
n_q}\frac{e^{ik|r-q|}}{|r-q|}] dS_q.
\end{equation*}
The integral above with $e^{ikr\cdot p_0}$ instead of $\psi^{eik}$ is equal to zero. Thus
\begin{equation*}
u^0\approx\frac{1}{4\pi}\int_{\partial \mathcal O}
[-\frac{\partial}{\partial n_q} (\Psi^{eik}(q))
\frac{e^{ik|r-q|}}{|r-q|} + \Psi^{eik}(q)\frac{\partial}{\partial
n_q}\frac{e^{ik|r-q|}}{|r-q|} ]dS_q.
\end{equation*}
The exact definition of $u^0$, in our case where $\partial \mathcal O$ is non-smooth, differs only by the introduction of a cut-off function $\eta=\eta(k,q)$ in the
 integrand above which makes the integrand smoother and provides important uniform estimates of $u^0$. Thus
\begin{equation}\label{appEik2}
u^0=\frac{1}{4\pi}\int_{\partial \mathcal O}
\eta[-\frac{\partial}{\partial n_q}( \Psi^{eik}(q))
\frac{e^{ik|r-q|}}{|r-q|}  + \Psi^{eik}(q)\frac{\partial}{\partial
n_q}\frac{e^{ik|r-q|}}{|r-q|} ]dS_q.
\end{equation}
Here $\eta$ is a $C^{\infty}$-function defined on the faces of the polyhedron $\partial
\mathcal O$ which vanishes in a $k^{-\delta}$-neighborhood of the edges of $\partial
\mathcal O$, equals one outside of a $2k^{-\delta}$-neighborhood of the edges, and is such that $|D^m\eta|<C_mk^{\delta m},~k>1,$ for each partial derivative $D^m$ of order $m$ on faces of $\partial
\mathcal O$. One can choose $\delta$ arbitrarily in the interval $0<\delta<1/2$. This choice of $\delta$ allows one
to apply the stationary phase method when an amplitude contains the factor $\eta$. However, some of the arguments at the
 end of section 7 become simpler when  $\delta$ is small enough. Indeed it is sufficient to fix $\delta=1/4$ and so
\begin{equation}\label{eta}
|D^m\eta|<C_mk^{m/4},~~k>1.
\end{equation}

The function $u^0$ satisfies (\ref{helm}), (\ref{Somm}). Furthermore, it will be proved in section 5 that the following
statement holds
\begin{lemma}\label{lemma0}
\begin{equation} \label{111}
\left \| \left . \left ( \frac{\partial}{\partial n} + k \lambda
\right ) \left ( u^0(r) -u(r) \right ) \right |_{\partial
\mathcal O} \right \|_{L_2(\partial \mathcal O)} =o(k), \quad k
\rightarrow \infty.
\end{equation}
\end{lemma}
This Lemma together with an \textit{a priori} estimate for the solutions of the problem (\ref{helm})-(\ref{bcOmega}) with $\Im \lambda>0$ obtained
in \cite{dtn}, allows one to justify the approximations $u\sim u^0,~ u_\infty \sim u^0_\infty,~k\rightarrow\infty. $
All the theorems in \cite{dtn} were formulated for quite smooth
obstacles. However, the proof of the Theorem 2 of that paper is
based only on an application of the Green formula, which is valid
for piecewise smooth obstacles also. Due to the importance  (and simplicity) of this theorem for our purposes, we prove
 a slightly improved version of this theorem here.
\begin{theorem}\label{th2a}
Let the function $v$ satisfy (\ref{helm}), (\ref{Somm}) and the boundary condition
\begin{equation*}
\left (\frac{\partial}{\partial n} + k \lambda \right )v = f, \quad \quad r=(x,y,z) \in
\partial \mathcal O,~~ \Im \lambda>0.
\end{equation*}
Then
\begin{equation}\label{apr}
|| v_{\infty}||_{L_2(S^2)} \leq \frac{1}{2k\sqrt{\Im
\lambda}} ||f||_{L_2(\partial \mathcal O)},~~|| v||_{L_2(\partial \mathcal O)}+||\frac{1}{k\lambda} \frac{\partial v}{\partial n}||_{L_2(\partial \mathcal O)}\leq \frac{1}{k\Im
\lambda} ||f||_{L_2(\partial \mathcal O)}.
\end{equation}
\end{theorem}
{\bf Proof.}  Taking the imaginary parts of both sides of the Green formula
\[
\int_{\Omega}(\Delta v+k^2 v)\overline{v}dx=\int_{\partial\Omega}v_n\overline{v}dS-ik\int_{S^2}|v_{\infty}|^2dS-\int_{\Omega}(|\nabla v|^2-k^2 |v|^2)dx,
\]
we obtain
$\Im \int_{\partial\Omega}v_n\overline{v}dS-k\int_{S^2}|v_{\infty}|^2dS=0.$
Thus, for any $c>0$,
\[
k\Im\lambda \int_{\partial\Omega}|v|^2dS+k\int_{S^2}|v_{\infty}|^2dS\leq \int_{\partial\Omega}|fv|dS
\leq ck\Im \lambda\int_{\partial\Omega}|v|^2dS+\frac{1}{4ck\Im\lambda}\int_{\partial\Omega}|f|^2dS.
\]
Choosing $c=1,$ or $c=1/2$ we arrive to the first inequality (\ref{apr}) or to the second inequality for $v$, respectively. The second inequality for $v_n$ follows from the boundary condition. The proof is complete.

Lemma \ref{lemma0} and Theorem \ref{th2a} imply
\begin{equation}\label{ResTh2}
\| u^0_\infty - u_\infty \|_{L_2(S^2)} = o(1), \quad k \rightarrow\infty,~~\Im \lambda>0.
\end{equation}
Evidently, for any $\varphi \in C(S^2)$ we have
\begin{equation}\label{DistrLimit}
\int_{S^2} \varphi(\theta) |u_\infty (\theta)|^2 dS_\theta = \int_{S^2} \varphi(\theta) |u^0_\infty (\theta)|^2 dS_\theta + o(1),
\quad k \rightarrow \infty,~~\Im \lambda>0.
\end{equation}
Hence, to prove Theorem \ref{tint} we need only prove Lemma \ref{lemma0}
and to analyze the far field behavior of $u^0$.

  Lemma  \ref{lemma0} and the far field behavior of $u^0$ will be derived from the following lemma on the near field
behavior of the Kirchhoff approximation. To state the lemma we need to introduce the set $\widehat{\mathcal
B}$  which is formed by extended to infinity lateral boundaries of shadow and reflected regions. One can define
$\widehat{\mathcal
B}$ as follows. Referring to Fig.~\ref{fig2}, let $\partial \mathcal O^1$ be
the upper part of $\partial \mathcal O$ which is struck by the
incident wave, and let $ \partial \mathcal O^2$ be the lower
 part of $\partial \mathcal O$ located strictly below $\partial \mathcal O^1$ (it is struck by the reflected wave). Their
left and right halves (rectangles) will be denoted by
 $\partial \mathcal O^{i}_{s}, i=1,2; s \in \{l,r\}$. Every time a geometrical optics ray strikes
 (directly or after a reflection) an edge of one of the four faces $\partial \mathcal O^{i}_{s}, i=1,2; s \in \{l,r\}$ of
 the obstacle $\mathcal O$, we extend the ray to infinity in both directions. That is in the direction of the incident ray
and in the direction of rays reflected from the face. These extensions form $\widehat{\mathcal B}$. Consider, for example
the incident ray $(GA'')$ on Fig.~\ref{fig2}. Then $\widehat{\mathcal B}$ contains half-infinite ray $(A''B)$ starting at $A''$ and
half-infinite ray starting at $B$ and propagating downwards.
\begin{lemma}\label{lemma0e}
The following asymptotics hold uniformly on any compact of $\partial \mathcal O$ which does not contain points of edges and
on any compact set in $\mathbb R^3 \backslash
\widehat{\mathcal B}:$
\begin{equation}\label{15}
u^0(r)=\psi^{eik}(r)+O(1/\sqrt k), \quad \left |\nabla u^0(r)-\nabla
\psi^{eik}(r) \right | = O(\sqrt k), \quad k\rightarrow \infty .
\end{equation}

For any $R>0$ there exist constants $C_i=C_i(R), i=1,2$ such that
\begin{equation}\label{16}
|u^0(r)-\psi^{eik}(r)|<C_1, \quad \left |\nabla u^0(r)-\nabla
\psi^{eik}(r) \right | < kC_2
\end{equation}
when $|r|<R, \quad k \rightarrow \infty$.
\end{lemma}

{\bf Remark.}
Note that $\Psi^{eik}$ has jumps on $\mathcal B$. In all estimates involving $\Psi^{eik}$ on $\partial \mathcal O$ or $\mathcal B$ we assume that one-sided limits
of the corresponding functions are considered.

 Lemmas \ref{lemma0e} and \ref{lemma0} will be proved in section 5. After that Theorem \ref{tint} follows from the following two well
known formulas:
\begin{equation}\label{FarFieldExpres}
u^0_\infty(\theta)=\frac{-1}{4\pi} \int_{Q} \left [\frac{\partial u^0}{\partial n}(r) + ik\left (\frac{r}{|r|} \cdot \theta \right )u^0 \right ] e^{-ik(\theta \cdot r)} dS, \quad \theta \in S^2,
\end{equation}
\begin{equation}\label{TotalCrossExpres}
\| u^0_\infty\|_{L_2(S^2)} = \frac{1}{k} \Im \int_{Q}  \frac{\partial u^0}{\partial n} \overline{u^0} dS,
\end{equation}
where $Q$ is a closed surface such  that the bounded part of the space with the boundary $Q$ contains
$\mathcal O. $ Note, one can take $Q=\partial \mathcal O$.

\section{Scattering by a polygon $M \subset \mathbb R^3$}
This section is devoted to a study of a simplified version of the problem under consideration. It concerns the scattering
by a single polygon which represents a typical face of the polyhedron $\mathcal O$. The obtained results
will be used later when the polyhedron $\mathcal O$ is considered.

Let $P \subset\mathbb R^3$ be a plane in $\mathbb R^3$, and let
$\alpha \in S^2$ be a unit vector which is transversal to $P$. We
choose a unit normal vector $n \in S^2$ to $P$ in such a way that
$ n_\alpha=\cos \theta = (n, \alpha) <0.$

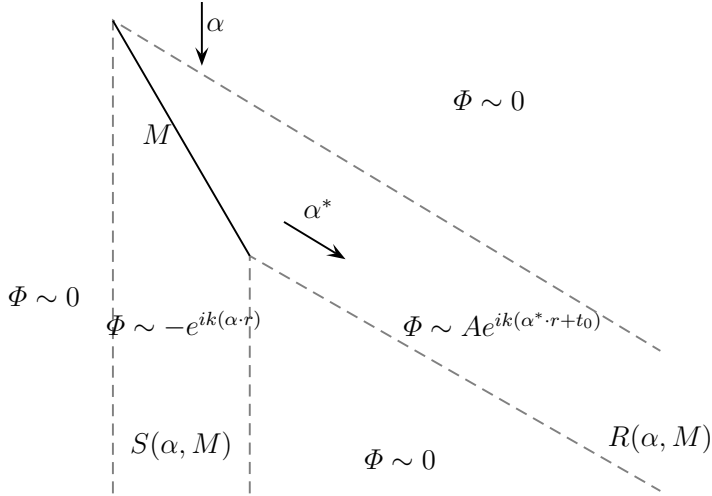
\begin{figure}[h]
\begin{picture}(0,180)
\scalebox{0.9}{ \rput(6.7,3.5){

\psline[linecolor=gray,linestyle=dashed](-2,-3.5)(-2,0)(4,-3.4641)

\rput(1.7,-1){$\varPhi \sim A e^{ik({\alpha^*} \cdot r+t_0)}$}

\rput(-5,-0.6){$\varPhi \sim 0$}
\rput(1.5,2.3){$\varPhi \sim 0$}
\rput(0.2,-3){$\varPhi \sim 0$}

 \rput(-3.35,1.8){$M$}

\psline[linecolor=gray,linestyle=dashed](-4,-3.5)(-4,3.47)(4,-1.4)

\pspolygon[fillstyle=solid,fillcolor=lightgray](-2,0)(-4,3.4641)(-2,0)

 \psline[arrows=->,arrowscale=1.8](-2.7,3.74)(-2.7,2.8)
 \rput(-2.5,3.4){$\alpha$}


\rput(-2.95,-1){$\varPhi  \sim -e^{ik(\alpha \cdot r)}$}

 \psline[arrows=->,arrowscale=1.8](-1.5,0.5)(-0.6,-0.05)
 \rput(-1.0,0.7){${\alpha^*}$}

 \rput(-3.0,-2.8){$S(\alpha,M)$}
 \rput(4.0,-2.7){$R(\alpha,M)$}

} }
\end{picture}
\caption{The figure presents the main term of the
asymptotics of the Kirchhoff approximation
$\varPhi=\varPhi_{\alpha,M}$ as $k\rightarrow\infty$. Here
$R(\alpha,M)$ is the reflected zone, $S(\alpha,M)$ is the shadow
zone, and $\varPhi$ vanishes outside of $\overline{R(\alpha,M) \bigcup S(\alpha,M)}$ when
$k\rightarrow\infty$. The main term of asymptotics coincides with
$\psi^{eik}_{\alpha,M}$. A rigorous statement is given by Lemma
\ref{EIKlemma0e}.}
\label{fig3}
\end{figure}

 Let $M$ be a convex
domain in $P$ bounded by a polygon $\partial M$. By the shadow
zone $\LL(\alpha)=\LL(\alpha,M) \subset \mathbb R^3$ we mean the
domain
which is inaccessible for the direct ray propagating along the vector
$\alpha$  assuming that $P \backslash M$ is transparent and that $M$
reflects the ray, i.e., $
\LL(\alpha) = \{ r=r_0 + t \alpha \in \mathbb R^3 ~  :
r_0 \in M, ~  t >0 \}.
$
We define the reflected zone to be the area
$\I(\alpha)=\I(\alpha,M) \subset \mathbb R^3$ covered by the
reflected rays, i.e.,
$\I(\alpha) = \{ r=r_0 + t \alpha^* \in \mathbb R^3 ~ : ~
r_0 \in M, ~ t >  0 \},$
where
$\alpha^*= \alpha^*(M)=\alpha - 2(\alpha \cdot n)n \in S^2$
is the reflection direction. In future $\alpha$ will be always equal to $p_0$ (see Fig.~\ref{fig3}) or $p_0^*$ (see Fig.~\ref{fig4}).

Denote by $D(r)=D_{\alpha,M}(r)$ the single layer potential with density
$\eta e^{ik (\alpha \cdot r)}:$
\begin{equation}\label{DDef}
D(r)=\int_{M} \eta e^{ik (\alpha \cdot q)}\frac{e^{ik|r-q|}}{|r-q|}
dS(q), \quad r \in \mathbb R^3,
\end{equation}
where $\eta$ is the function introduced in (\ref{appEik2}). For any $r \in  M$, define $r_\varepsilon = r + n \varepsilon$,
 $\varepsilon>0$ and let
\begin{equation}\label{dnD}
\frac{\partial}{\partial n} D(r)=\lim_{\varepsilon \rightarrow
0}\frac{\partial}{\partial n} D(r_\varepsilon).
\end{equation}

Denote by  $N(r)=N_{\alpha M}(r)$ the double layer  potential with density $\eta e^{ik (\alpha \cdot r)}:$
\begin{equation}\label{NDef}
N(r)=\int_{M}\eta e^{ik (\alpha \cdot q)}\frac{\partial}{\partial n_q}
\frac{e^{ik|r-q|}}{|r-q|} dS(q), \quad q \not \in M,
\end{equation}
and define its value on $M$ similar to (\ref{dnD}), i.e.
\begin{equation}\label{dnN}
 \quad N(r)=\lim_{\varepsilon \rightarrow 0}
 N(r_\varepsilon),~~~\frac{\partial}{\partial n} N(r)=\lim_{\varepsilon \rightarrow 0}\frac{\partial}{\partial n}
 N(r_\varepsilon),~~r \in M.
\end{equation}
Finally define  $A= A(\alpha,M)=\frac{i(n \cdot
\alpha)+\lambda}{i(n \cdot \alpha)-\lambda}$ (see (\ref{factor})).

 Consider scattering
of $e^{ikr\cdot \alpha}$ by the obstacle consisting of the surface
$M$ only, and define the eikonal approximation
$\Psi^{eik}_{\alpha,M}$ as in (\ref{app1im}).
 Thus (see Fig.~\ref{fig3})
$$
\psi^{eik}_{\alpha,M}(r) := \Psi^{eik}_{\alpha,M}(r) -e^{ikr\cdot \alpha} = \left \{
\begin{array}{cc}
-e^{ikr\cdot \alpha}, & r \in \LL(\alpha,M) \\
A e^{ik[(\alpha^* \cdot r)+t_0]}, & r \in \I(\alpha,M) \\
0, & \mathbb R^3 \backslash \{ \I(\alpha) \cup \LL(\alpha) \}
\end{array}
\right .
$$
where the constant $t_0=t_0(\alpha,M)$ is defined by the relation
\begin{equation}\label{EIKCTransmission}
e^{ik[(\alpha^* \cdot r)+t_0]}=e^{ik(\alpha \cdot r)},~ \quad r \in
M.
\end{equation}
Since $\alpha^*=\alpha-2(\alpha \cdot n)n$, (\ref{EIKCTransmission}) implies that
\begin{equation}\label{t0}
t_0=2(\alpha \cdot n)(n \cdot r),~ \quad r \in
M.
\end{equation}
Note that $n \cdot (r_1-r_2)=0$ when $r_1,r_2 \in
M.$ Thus, $t_0$ is a constant.

 The function
$\Psi^{eik}_{\alpha,M}(r)$  has one-sided limiting values on both sides of $M$, the value of the function and of its normal
derivative is zero on the shadow side of $M$. On the illuminated side $M_+$ of $M$:
\begin{equation}\label{psiek}
\Psi^{eik}_{\alpha,M}(r)=(A+1)e^{ikr\cdot \alpha},~~
\frac{\partial}{\partial n}\Psi^{eik}_{\alpha,M}(r)=ik(-A+1)n_{\alpha}e^{ikr\cdot \alpha},~~~r \in M_+.
\end{equation}
Thus the Kirchhoff approximation (\ref{appEik2}) in the case of
$M=\mathcal O$  takes the form
\begin{equation}\label{impedanceCompostionDef}
\varPhi=\varPhi_{\alpha,M}=\frac{ik}{4\pi} (A-1) n_{\alpha}
D_{\alpha,M} + \frac{A+1}{4\pi} N_{\alpha, M}.
\end{equation}

The following lemma plays a crucial role in the proof of the main
result, theorem \ref{tint}. Informally it states that the Kirchhoff
approximation in the case of $\mathcal O=M$ is close to
$\psi^{eik}_{\alpha,M} := \Psi^{eik}_{\alpha,M}-e^{ikr\cdot
\alpha}$ when $k\gg 1,$ and it justifies Fig.~\ref{fig3}.
\begin{lemma}\label{EIKlemma0e}
The following asymptotics as $k \rightarrow \infty$ hold uniformly on any compact set in $M$ and on any compact set of
$\mathbb R^3 \backslash \overline{R(\alpha,M) \bigcup S(\alpha,M)}$
\begin{equation} \label{1}
\varPhi_{\alpha,M}(r)= \psi^{eik}_{\alpha,M}(r)+O(1/\sqrt k), \quad \left
|\nabla \varPhi_{\alpha,M}(r)-\nabla \psi^{eik}_{\alpha,M}(r)
\right | = O(\sqrt k).
\end{equation}
For any $R>0$ there exist constants $C_i=C_i(R), i=1,2$ such that
\begin{equation}\label{2}
|\varPhi_{\alpha,M}(r)-\psi^{eik}_{\alpha,M}(r)|<C_1, \quad \left
|\nabla \varPhi_{\alpha,M}(r)-\nabla \psi^{eik}_{\alpha,M}(r)
\right | < kC_2,
\end{equation}
when $|r|<R, \, k \rightarrow \infty$.
\end{lemma}

{\bf Remarks. 1)}
Note that $\Psi^{eik}_{\alpha,M}$ has jump discontinuities on the boundaries of the reflected and shadow zones. In all estimates involving $\Psi^{eik}$ on $M$ or on the lateral sides of the boundaries of the reflected and shadow zones, we assume that one-sided limits
of the corresponding functions are considered.

{\bf 2)} Since $\psi_{\alpha,M}$ satisfies the impedance boundary
condition (\ref{bcOmega}), the estimates (\ref{1}), (\ref{2}) imply
that $$ \frac{\partial}{\partial n} \varPhi_{\alpha,M} +
k\lambda \varPhi_{\alpha,M}=-k (i(n \cdot \alpha) + \lambda)
e^{ik(\alpha \cdot r)}+  O(\sqrt k),~ k \rightarrow \infty,
$$
where the estimate of the remainder is uniform on any compact subset of  $M$, and is bounded by $Ck$ on $M$.

The proof of Lemma \ref{EIKlemma0e} is rather technical and is given in
Appendix 1.

\section{Asymptotics of the Kirchhoff approximation in a bounded region}

\textbf{Proof of Lemma \ref{lemma0e}}.
Recall, see Fig.~\ref{fig2}, that $\partial \mathcal O^{1}_{l}, \partial \mathcal O^{1}_{r}$ are the upper
parts of the $\partial \mathcal O$ which are struck by the incident wave, and that  $\partial \mathcal O^{2}_{l}, \partial \mathcal O^{2}_{r}$ are the lower
 parts of the $\partial \mathcal O$ which are struck by the reflected wave.

Note that function $\Psi^{eik}$ in (\ref{appEik2}) vanishes on the
part of the boundary of the obstacle which is not accessible for
the rays, i.e., $\partial\mathcal O$ in (\ref{appEik2}) can be
replaced by the union of four polygons $\partial
O^i_j,~i=1,2,j=l,r$. We split them in two pairs where the
polygons in each pair are connected by rays. The first pair
consists of $M_1= \partial O^1_l$ and $ M_2= \partial O^2_r$. Then
$u^0$ can be written in the form
\begin{equation}\label{klr}
u^0=\mathcal L+\mathcal R,
\end{equation}
where
\begin{equation}\label{u1}
\mathcal L=-\frac{1}{4\pi}\int_{M_1\bigcup
M_2}\eta
\frac{\partial}{\partial n_q} (\Psi^{eik}(q))
\frac{e^{ik|r-q|}}{|r-q|} dS_q
+ \frac{1}{4\pi}\int_{M_1\bigcup
M_2}
 \eta\Psi^{eik}(q)\frac{\partial}{\partial
n_q}\frac{e^{ik|r-q|}}{|r-q|} dS_q,
\end{equation}
and $\mathcal R$ is given by a similar expression with $\partial
O^1_l\bigcup\partial O^2_r$ in place of $M _1\bigcup
M_2$.
\begin{figure}[h]
\begin{picture}(0,170)
\scalebox{0.8}{ \rput(6.7,3.8){

\psline[linecolor=gray,linestyle=dashed](-2,-3.5)(-2,0)(4,-3.4641)

\rput(1.7,-1){$\varPhi_{2,r} \sim -A e^{ik({\alpha^*} \cdot r+t_0^1)}$}

\rput(-5,-1){$\varPhi_{2,r} \sim 0$}
\rput(1.5,2.3){$\varPhi_{2,r} \sim 0$}
\rput(0.2,-3){$\varPhi_{2,r} \sim 0$}

 \rput(-3.35,1.8){$M_2$}

\psline[linecolor=gray,linestyle=dashed](-4,-3.5)(-4,3.47)(4,-1.4)

\pspolygon[fillstyle=solid,fillcolor=lightgray](-2,0)(-4,3.4641)(-2,0)

 \psline[arrows=->,arrowscale=1.8](-3.4,1.04)(-3.4,0.0)
 \rput(-3.2,0.54){$\alpha^*=p_0$}

\rput(-2.8,-2){$\varPhi_{2,r}  \sim A^2e^{ik(\alpha^* \cdot r)+t_0^1+t_0^2}$}

 \psline[arrows=->,arrowscale=1.8](-5.5,3.5)(-4.6,2.95)
 \rput(-5.0,3.7){${\alpha=p_0^*}$}


} }
\end{picture}
 \caption{The main term of asymptotics of $\varPhi_{2,r}$}
\label{fig4}
\end{figure}
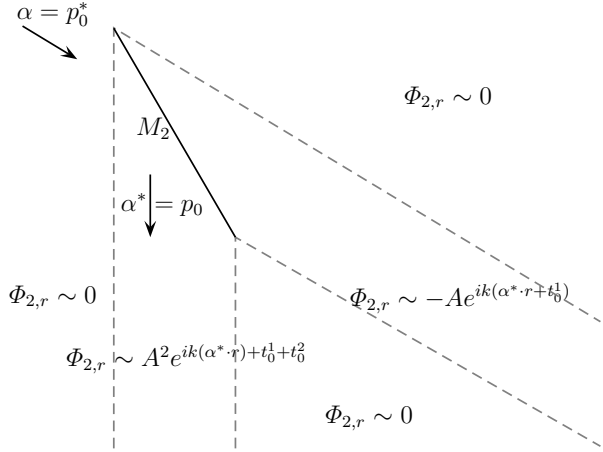
We represent $\mathcal L$ as $\mathcal
L=\varPhi_{1,l}+\varPhi_{2,r}$ where the first term in the sum
involves integration only over $M_1$ and the second term involves
integration only over $M_2$. Then $\varPhi_{1,l}$ coincides with
the Kirchhoff approximation $u^0$ for a single polygon which is given by (\ref{impedanceCompostionDef}) as studied in
the previous section. In particular, Fig.~\ref{fig3} and Lemma
\ref{EIKlemma0e} with $\alpha=p_0$ are valid for $\varPhi_{1,l}$.

\begin{figure}[h]
\begin{picture}(0,155)
\scalebox{0.75}{ \rput(6.7,4.2){

\psline[linecolor=gray,linestyle=dashed](-2,-4.4641)(-2,0)(4,-3.4641)(4,-5)

 \rput(-3.35,1.7){$M_1$}
\rput(3.3,-1.4){$M_2$}

\psline[linecolor=gray,linestyle=dashed](-4,3.47)(2,0)(2,-4)

\psline[linecolor=gray,linestyle=dashed](-4,3.4641)(-4,-4.0)

\pspolygon[fillstyle=solid,fillcolor=lightgray](-2,0)(-4,3.4641)(-2,0)
\pspolygon[fillstyle=solid,fillcolor=lightgray](2,0)(4,-3.4641)(2,0)


\rput(-2.8,-2){$-e^{ik({p_0} \cdot r)}$}

 \psline[arrows=->,arrowscale=1.8](-1.5,1.5)(-0.6,0.95)
 \rput(-1.0,0.7){${p_0^*}$}

\rput(0.6,0){$ A e^{ik({p_0^*} \cdot r+t_0^1)}$} \rput(2.9,-4){$
A^2 e^{ik({p_0} \cdot r+t_0^1+t_0^2)}$}

} }
\end{picture}
\caption{The main term of asymptotics of $\mathcal L$}
\label{fig5}
\end{figure}
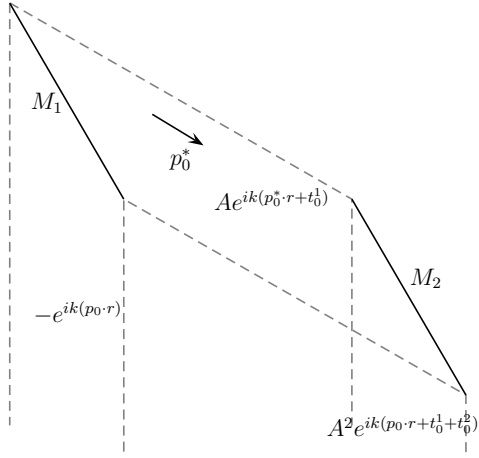

Function
$\Psi^{eik}$ in the region covered by the rays between $M_1$ and
$M_2$ has the  form of a plane wave propagating in the direction
$p_0^*$, i.e., $\varPhi_{2,r}$ is also a Kirchhoff approximation
for a single polygon. The only difference between $\varPhi_{1,r}$ and $\varPhi_{2,r}$ is
that in the second case the direction $\alpha$ of the incident wave is
different ($\alpha=p_0^*$) and the incident wave contains a constant factor
$Ae^{ikt_0}.$ Thus, the main term of asymptotics of $\varPhi_{2,r}$ is given by Fig.~\ref{fig4}.
This figure coincides with Fig.~\ref{fig3} with $\alpha=p^*_0$ and extra factor $Ae^{ikt_0}$ added.
An analogue of Lemma \ref{EIKlemma0e} is valid for $\varPhi_{2,r}$ stating that  $\varPhi_{2,r}$ differs from the main term
indicated in Fig.~\ref{fig4} by $O(1/\sqrt k)$ on any compact outside of the lateral sides of reflected and shadow zones and
by $O(1)$ on any compact set.
The remainder is multiplied by $k$ after differentiation. Thus, the main term of $\mathcal L =  \varPhi_{1,l} + \varPhi_{2,r}$
is indicated in Fig.~\ref{fig5}. The shift of the phase $t_0^1+t_0^2$ in Fig.~\ref{fig5} was discussed in section 2, where it was noted that this shift is equal to $\Delta$ (see Fig.~\ref{fig2}). Since $\mathcal R$ has a similar asymptotic expansion and
$u^0=\mathcal L + \mathcal R$, this proves the statements of Lemma \ref{lemma0e}.

\textbf{Proof of Lemma \ref{lemma0}}. From (\ref{bcOmega}) and (\ref{factor}) it follows that
\begin{equation}\label{bcOm}
\left (\frac{\partial}{\partial n} + k \lambda \right )(u-\psi^{eik}(r)) = 0, \quad \quad r \in
\partial \mathcal O.
\end{equation}
On the other hand, Lemma \ref{lemma0e} shows that $u^0$ is close to $\psi^{eik}(r)$ near $\partial \mathcal O$. This leads to the statement of Lemma \ref{lemma0}. Indeed, let us fix an arbitrary $\varepsilon>0$ and let $\partial \mathcal O_{\varepsilon}$ be such a small neighborhood of the edges on $\partial \mathcal O$ that the area  $|\partial \mathcal O_{\varepsilon}|$ of that neighborhood does not exceed
$ \frac{\varepsilon^2}{2(C_1|\lambda|+C_2)^2},$
 where $C_1,C_2$ are constants defined in (\ref{16}). Then, for any $k>0$,
\begin{equation}\label{a1}
\int_{\partial \mathcal O_{\varepsilon}} |(\frac{\partial}{\partial n} + k \lambda )(u^0-\psi^{eik}(r))|^2dS<(\varepsilon k)^2/2.
\end{equation}
Furthermore, (\ref{15}) implies that
\begin{equation}\label{a2}
\int_{\partial \mathcal O\backslash\partial \mathcal O_{\varepsilon}} |(\frac{\partial}{\partial n} + k \lambda )(u^0-\psi^{eik}(r))|^2dS<Ck<(\varepsilon k)^2/2
\end{equation}
if $k$ is large enough. Statement of Lemma \ref{lemma0} follows immediately from (\ref{bcOm})-(\ref{a2}).
\section{Asymptotic behavior of the total cross section}
{\bf Proof of the second statement of Theorem \ref{tint}}. We begin with the evaluation of the total cross section of the
 Kirchhoff approximation $u^0$ using (\ref{TotalCrossExpres}). Consider the boundary
$Q$ of a cube such that its faces are parallel to the coordinate planes and which contains the obstacle $\mathcal O$. For
 any small $\delta >0$, we split $Q$ in the following three parts: $Q=Q_1\bigcup Q_2 \bigcup Q_3$, where $Q_1$ is a
 $\delta$-neighborhood of $Q\bigcap \widehat{\mathcal B}$ (see definition of $\widehat{\mathcal B}$ in Lemma
\ref{lemma0e}), $Q_2$ is the orthogonal projection of the obstacle $\mathcal O$ into the low horizontal side of $Q$
without points already included into $Q_1$, and $Q_3$ is the remaining part of $Q$.

Note that area $|Q_2|$ of $Q_2$ converges to $1/2$ as $\delta\rightarrow 0$.
 Thus, from (\ref{16}) it follows that for any $\varepsilon>0$ one can choose $\delta$ such that
\begin{equation}\label{q2}
\frac{1}{k} \int_{Q_1}| \overline{u}^0 \frac{\partial u^0}{\partial r}|dS\leq\frac{\varepsilon}{4}~~\text{and} ~0<\frac{1}{2}-|Q_2|\leq\frac{\varepsilon}{4}.
\end{equation}
With $\delta$ fixed, one can choose $k_0$ such that
\begin{equation}\label{q}
\frac{1}{k} \int_{Q_3}| \overline{u}^0 \frac{\partial u^0}{\partial r}|dS\leq\frac{\varepsilon}{4},~~
\frac{1}{k} |\Im \int_{Q_2} \overline{u}^0 \frac{\partial u^0}{\partial r}dS-Z|\leq\frac{\varepsilon}{4},
\end{equation}
where $Z=|-1+A^2e^{ik\Delta} |^2$.
The latter relations follow from (\ref{15}), since (see Fig.~\ref{fig2}) $\psi^{eik}=\Psi^{eik}-e^{ikz}=0$ on $Q_3$, and
\[
\psi^{eik}=A^2e^{ik(z+\Delta)}-e^{ikz}~~~\text{on}~~Q_1.
\]

Inequalities (\ref{q2}), (\ref{q}) and formula (\ref{TotalCrossExpres}) imply the validity of (\ref{st2}) for $u^0$. Its validity for $u$ follows from (\ref{ResTh2}).
 The proof is complete.

\section{Asymptotics of the transport cross section}

The goal of this section is to show that the transport cross section for the obstacle under consideration vanishes as
 $k\rightarrow\infty$ if $\Im \lambda>0$. The last statement of theorem \ref{tint} follows immediately (see the last paragraph of
 this section) from the first two statements. As the first step, we need to obtain an integral representation for the
scattering amplitude of the Kirchhoff approximation. Recall, that
$u^0=\mathcal L+\mathcal R$ (see (\ref{klr})), where $\mathcal L$ is given by (\ref{u1}) and $\mathcal R$ has a similar
 form. Denote the scattering amplitude of the functions $u^0,~\mathcal L,~\mathcal R$ by
 $u_{\infty}^0,~\mathcal L_{\infty},~\mathcal R_{\infty}$, respectively (see (\ref{scamp}) where the scattering amplitude
 is defined for $u$). Recall that $M_1= \partial O^1_l,~M_2= \partial O^2_r$, and $M_2$ is the shift of $M_1$; to be more
 precise, $M_2=M_ 1+dp_0^*,~d>0$.
\begin{lemma}\label{FarF}
We have
\begin{equation}\label{ulr}
u_{\infty}^0=\mathcal L_{\infty}+\mathcal R_{\infty},
\end{equation}
where
\begin{equation}\label{u1inf2}
\mathcal L_{\infty}=\frac{ik}{4\pi}\Psi(\theta)\int_{M_1}\eta e^{ik(p_0-\theta)\cdot q} dS_q
\end{equation}
with
\begin{equation}\label{psitet}
\Psi(\theta)=  (A-1)n_{p_0} -(A+1)n_{\theta} +
 A e^{ikd(1-p_0^* \cdot \theta)}  \left ( (A-1)n_{p_0} + (A+1)n_{\theta}\right ).
\end{equation}
Function $\mathcal R_{\infty}$ is the reflection of $\mathcal L_{\infty}$ with respect to the first argument, i.e $\mathcal R_{\infty}(\theta)=\mathcal L_{\infty}(\overline{\theta})$, where $\overline{\theta}=\overline{\theta}(\theta)=(-\theta_1,\theta_2,\theta_3)$.

\end{lemma}

\textbf{Proof.} The validity of (\ref{ulr}) follows from (\ref{klr}), so one needs only to find the terms in the right hand
side of (\ref{ulr}). Scattering amplitude $\mathcal L_{\infty}$ can be found from  (\ref{u1}):
$$
\mathcal L_{\infty}=-\frac{1}{4\pi}\int_{M_1\cup M_2}\eta
\frac{\partial}{\partial n_q} (\Psi^{eik}(q))
e^{-ik\theta\cdot q} dS_q - \frac{ikn_\theta}{4\pi}\int_{M_1\cup M_2}
 \eta\Psi^{eik}(q)e^{-ik\theta\cdot q} dS_q
$$
\begin{equation}\label{u1inf}
=-\frac{1}{4\pi}\int_{M_{1}\cup M_2}\eta(\frac{\partial}{\partial n_q} (\Psi^{eik}(q))+ikn_\theta\Psi^{eik}(q))e^{-ik\theta\cdot q} dS_q,~~\theta \in S^2.
\end{equation}

We need to insert here the values of the eikonal approximation $\Psi^{eik}(q)$ on $M_1$ and $M_2.$ Its value and the
 value of the normal derivative on $M_1$ are given by (\ref{psiek}) with $\alpha=p_0$. In order to find the corresponding values
on $M_2$ one needs to take into account the following fact which was discussed in section 5. The wave $\Psi^{eik}(r)$
shortly before the incident to $M_2$ has the form $Ae^{ik(p_0^*\cdot r+t_0)}$, i.e., it differs from the original
wave $e^{ikp_0\cdot r}$ coming to $M_1$ by the choice of the incident direction ($p_0^*$ instead of $p_0$) and an extra
 factor $Ae^{ikt_0}$. Thus,  $\Psi^{eik}(r)$ and its derivative on $M_2$ are given by (\ref{psiek}) with $\alpha$ replaced
by $p_0^*$ and the extra factor $Ae^{ikt_0}$ added in the right-hand sides of (\ref{psiek}). Hence, (\ref{u1inf}) can be
 rewritten in the form
\begin{eqnarray*}
\mathcal L_{\infty}=-\frac{ik}{4\pi}\int_{M_1}\eta((-A+1)n_{p_0}+(A+1)n_{\theta})e^{ik(p_0-\theta)\cdot q} dS_q
\\
-\frac{ikAe^{ikt_0}}{4\pi}\int_{M_2}\eta((-A+1)n_{p_0^*}+(A+1)n_{\theta})e^{ik(p_0^*-\theta)\cdot q} dS_q.
\end{eqnarray*}
In order to prove (\ref{u1inf2}), it remains only to rewrite the last integral above as an integral over $M_1$
by using the substitution $q\rightarrow q+dp_0^*$ and the following two facts.
Firstly, the normals $n$ on $M_1$ and $M_2$ have different direction, and therefore,
$n_{p_0^*}, n_{\theta}$ on $M_2$ are equal to $n_{p_0}, -n_{\theta}$ on $M_1$, respectively.
Secondly, since $p_0^*=p_0-2(p_0 \cdot n)n$ and $t_0$ is given by (\ref{t0}) where $M=M_1$,
the exponent in the second integral above (with $q$ replaced by $q+dp_0^*,~ q\in M_1$) can be rewritten as follows
$$
(p_0^* - \theta) \cdot (q+dp_0^*))=(p_0^* - \theta) \cdot q+d(1-p_0^* \cdot \theta)
$$
$$
=
(p_0-\theta)\cdot q - 2(n \cdot p_0) (n \cdot
q) +d(1-p_0^* \cdot \theta)=(p_0-\theta)\cdot q  +d(1-p_0^* \cdot \theta)-t_0.
$$
The symmetry between $\mathcal L_{\infty}$ and $\mathcal R_{\infty}$ can be observed from formula (\ref{u1inf}) for $\mathcal L_{\infty}$, and the corresponding formula for $\mathcal R_{\infty}$.

The proof of Lemma \ref{FarF} is complete.

\begin{theorem}\label{con1}
If $\Im \lambda>0$ then the transport cross section of the Kirchhoff approximation $u_0$ vanishes as $k\rightarrow\infty$.
 To be precise,
\[
\int_{S^2} (1-\theta \cdot p_0))|u^0_\infty(\theta)|^2
d\mu (\theta)=O(k^{-\varepsilon}),~~\varepsilon >0,~~k\rightarrow\infty.
\]
\end{theorem}
\textbf{Proof.} Denote the integral in (\ref{u1inf2}) by $D_{\infty}(\theta)$; it is equal to the scattering amplitude of
the single layer (\ref{DDef}). From Lemma \ref{FarF} it follows that it is enough to prove
 the statement of the theorem for the part $\mathcal L_{\infty}$ of $u_0$, i.e. Theorem \ref{con1} will be proved as soon
 as we show that
\begin{equation}\label{con11}
\int_{S^2}\beta_{p_0}(\theta)|\Psi(\theta)D_{\infty}(\theta)|^2dS=O(k^{-2-\varepsilon}),~~\varepsilon>0;
~~~\beta_{p_0}(\theta)=1-p_0\cdot\theta,~~~\Im\lambda>0.
\end{equation}

We do not need the exact form of the functions $\beta_{p_0},\Psi$, but only some estimates valid for those functions. Namely,
\begin{equation}\label{beta}
0\leq \beta_{p_0}(\theta) \leq C|\theta-p_0|^2,~~\theta \in S^2,
\end{equation}
\begin{equation}\label{psibeta}
|\Psi(\theta)|\leq C,~~|\Psi(\theta)|\leq C(|\theta-p^*_0|+k|\theta-p^*_0|^2),~~\theta \in S^2,
\end{equation}

In fact, $\beta$ is a real valued infinitely smooth function on $S^2$ with the minimum value at the point $\theta=p_0$. Thus, the validity of the quadratic estimate (\ref{beta}) is obvious. In order to justify (\ref{psibeta}), we represent $\Psi$ in the form $\Psi(\theta)=\Psi_1(\theta)+\Psi_2(\theta)$ where $\Psi_1(\theta)$ is given by (\ref{psitet}) with the exponential factor in that formula omitted, and
\[
\Psi_2(\theta)=\Psi(\theta)-\Psi_1(\theta)=A (e^{ikd(1-p_0^* \cdot \theta)}-1)  \left ( (A-1)n_{p_0} + (A+1)n_{\theta}\right ).
\]
Since $n_{p_0^*}=-n_{p_0}$, function $\Psi_1(\theta)$ vanishes at $\theta=p_0^*$. This function is infinitely smooth, complex-valued and $k$-independent. Thus, $|\Psi_1(\theta)|\leq C|\theta-p_0^*|$. Further,
\[
|\Psi_2(\theta)|\leq C|e^{ik\beta_{p_0^*}(\theta)}-1|\leq Ck|\theta-p_0^*|^2.
\]
The latter inequality follows immediately from (\ref{beta}). Hence, the validity of both (\ref{beta}) and (\ref{psibeta}) are justified.

For any vector $u\in S^2$ denote by $\widetilde{u}$ the orthogonal projection of $u$ on the plane $P$ containing $M_1$.
Vectors $p_0,p_0^*$ have the same projections $\widetilde{p_0}=\widetilde{p_0^*}$. Let $B_{\widetilde{p_0}}(k^{-\gamma})$
be the disk
in $P$ of the radius $k^{-\gamma},~\frac{2}{3}<\gamma<\frac{3}{4},$ centered at $\widetilde{p_0}$. Denote by
 $\Omega_1,\Omega_2$ small neighborhoods on $S^2$ of  the points $p_0,p_0^*$, respectively, whose projections on $P$
coincide with $B_{\widetilde{p_0}}(k^{-\gamma})$. Then
\begin{equation}\label{proj1}
|\widetilde{\theta}-\widetilde{p_0}|>k^{-\gamma}>0 ~~\text{when}~\theta \in \Omega_3=S^2\setminus(\Omega_1\cup\Omega_2).
\end{equation}
Denote by $\widetilde{\Delta}=\widetilde{\Delta}_p$ the two-dimensional Laplace operator in the plane $P$. For any
 $q,q_0\in M_1$ and $p=q-q_0\in P$, we have
\[
\widetilde{\Delta}_pe^{ik(p_0-\theta)\cdot q}=\widetilde{\Delta}_pe^{ik[(p_0-\theta)\cdot q_0+(\widetilde{p_0}-\widetilde{\theta})\cdot p]}=-k^2|\widetilde{p_0}-\widetilde{\theta}|^2e^{ik(p_0-\theta)\cdot q}.
\]
Hence, from (\ref{proj1}) and (\ref{eta}) it follows that for any $m$
\[
\mathcal D_{\infty}(\theta)=
\frac{1}{(ik)^{2m}|\widetilde{p_0}-\widetilde{\theta}|^{2m}}\int_{M_1}\eta \widetilde{\Delta}^m e^{ik(p_0-\theta)\cdot q} dS_q
\]
\[
=\frac{1}{(ik)^{2m}|\widetilde{p_0}-\widetilde{\theta}|^{2m}}\int_{M_1} (\widetilde{\Delta}^m\eta) e^{ik(p_0-\theta)\cdot q} dS_q=O(k^{-2m(\frac{3}{4}-\gamma)}),~~\theta \in \Omega_3.
\]
Since $\gamma<3/4$ and $m>0$ is arbitrary, it remains to show that
\begin{equation}\label{con111}
I_i:=k^2\int_{\Omega_i}\beta_{p_0}(\theta)|\Psi(\theta)D_{\infty}(\theta)|^2dS=O(k^{-\varepsilon}),
~~~i=1,2,~~~\Im\lambda>0.
\end{equation}

Let us prove (\ref{con111}) for $i=2$. The case $i=1$ can be treated similarly (in fact, the latter case  is simpler
with the reference to (\ref{beta}) instead of (\ref{psibeta})). We use Euclidean coordinates on $P$ as local coordinates
on $\Omega_2:~\theta=\theta(\widetilde{\theta})$. Since $p_0,p_0^* $ are transversal to $P$,
the Jacobian $J(\theta)=\frac{dS}{dS_{\widetilde{\theta}}}=1/\sqrt{(1-|\widetilde{\theta}|^2)}$ is bounded when $\theta\in \Omega_1,$ and
\[
|\widetilde{\theta}-\widetilde{p_0^*}|\leq |\theta-p_0^*|\leq C|\widetilde{\theta}-\widetilde{p_0^*}|,~~\theta\in \Omega_1.
\]
From here,  (\ref{psibeta}) and the boundedness of $\beta(\theta)$ and $|D_{\infty}(\theta)|$ it follows that
\begin{eqnarray*}
I_2\leq Ck^2\int_{B_{p_0}(k^{-\gamma})}|\Psi(\theta)|^2dS_{\widetilde{\theta}}\leq C\int_{B_{p_0}(k^{-\gamma})}(k^2|\widetilde{\theta}-\widetilde{p_0^*}|^2
+k^4|\widetilde{\theta}-\widetilde{p_0^*}|^4)dS_{\widetilde{\theta}}
\\
\leq 2\pi C\int_0^{k^{-\gamma}}(k^2\sigma^3+k^4\sigma^5)d\sigma=O(k^{-\varepsilon}),~~\varepsilon=6\gamma-4>0.
\end{eqnarray*}

The proof of Theorem \ref{con1} is complete.

{\bf Proof of the first and the last statements of Theorem
\ref{tint}}. The first statement of Theorem \ref{tint} follows
immediately from (\ref{DistrLimit}) and Theorem \ref{con1}. Let us
prove the last statement. It follows from the first statement that
for any neighborhood $S_{\delta}\subset S^2$ of the point $p_0$,
we have $ \int_{S^2\setminus
S_{\delta}}|u_{\infty}|^2dS\rightarrow 0,~~k\rightarrow\infty.$
 Then from the second statement we obtain that
\[
\int_{ S_{\delta}}|u_{\infty}|^2dS\rightarrow \frac{1}{2} \left |-1+A^2e^{ik\Delta} \right |^2,~~k\rightarrow\infty.
\]
Obviously, the latter two relations imply (\ref{DistrLimitO}).
\section{Real impedance}

{\bf Proof of Theorem \ref{tint1}.} Let us fix $\lambda_0 \in \mathbb R$. We will need the optical theorem
for $\lambda \in \mathbb R$:
\begin{equation}\label{212}
\sigma_{\lambda}(k)=(4\pi /k) \Im u_\infty(p_0), \quad \lambda \in
\mathbb R.
\end{equation}

Note that this theorem concerns the exact solution of the scattering problem, i.e., it is valid for $u_\infty$ but not for
the far field given by the Kirchhoff approximation $u_{\infty}^{0}$. Hence, asymptotic behavior
of $\Im u^0_\infty(p_0)$ can not be derived from the known
asymptotics of $\|u^0_\infty\|$. We therefore calculate it
independently. From lemma \ref{FarF} we have
$$
u^0_\infty(p_0) = \frac{ik}{2\pi} \Psi(p_0)\int_{M}\eta(q)dS_q =
\frac{ik}{\pi} n_{p_0}(-1+A^2e^{ikd(1-p_0\cdot
p_0^*)})\int_{M}\eta(q)dS_q.
$$
Note that $d=|A'B''|$ (see Fig.~\ref{fig2}) is the hypotenuse of the triangle $A'B''H$. Thus
$d(1-p_0\cdot p_0^*) = d - |B''H|=\Delta$, and
$$
u^0_\infty(p_0) = \frac{ik}{4\pi} (-1+A^2e^{ik\Delta})(1+o(1)),
\quad k \rightarrow \infty.
$$
Hence, if $k=k_n$ defined by (\ref{kfrom}) and $\lambda$ is real, then on noting that $A(\lambda)$ does not depend on $k$
and $|A|=1$ for
real $\lambda$,
\begin{equation}\label{auxRI1}
u^0_\infty(p_0)= \frac{ik_n}{4\pi}
(-1+e^{2i(Arg(A(\lambda))-Arg(A(\lambda_0))})(1+o(1)), \quad k_n
\rightarrow \infty,~~\Im\lambda=0.
\end{equation}

Our next goal is to estimate the difference $f_\lambda(p_0):=u^0_\infty(p_0)-u_\infty(p_0)$ for complex $\lambda,~~\Im\lambda>0$. First of all note that the relation (\ref{FarFieldExpres}) for the scattering amplitude is valid also for $u$ and $v=u^0-u$. Thus,
\[
|f_\lambda(p_0)|\leq || v_n||_{L_2(\partial \mathcal O)}+||k v||_{L_2(\partial \mathcal O)}.
\]
Denote by $\Lambda$ the set $\Lambda=\{\lambda:~|\lambda-\lambda_0|\leq 1,~\Im\lambda > 0\}.$ Statements of Lemma \ref{lemma0} and Theorem \ref{th2a} are valid uniformly in $\lambda \in \Lambda$. Together with the estimate above this implies that
\begin{equation}\label{ineqRI}
|f_\lambda(p_0)| \leq \frac{o(k)}{\Im \lambda}, \quad k
\rightarrow \infty,~~\lambda \in \Lambda,
\end{equation}
where $o(k)$ is uniform in $\lambda\in \Lambda$.

Let $\varepsilon > 0$. Note that $ \min_{\lambda \in
(\lambda_0-\varepsilon,\lambda_0+\varepsilon)}
(\lambda-(\lambda_0-2\varepsilon))((\lambda_0+2\varepsilon)-\lambda)=3\varepsilon^2.$
Thus,
\begin{equation}\label{estimin}
\varPhi(\lambda):=\frac{(\lambda-(\lambda_0-2\varepsilon))((\lambda_0+2\varepsilon)-\lambda)}{3\varepsilon^2}
\geq 1 , \quad \lambda \in
[\lambda_0-\varepsilon,\lambda_0+\varepsilon].
\end{equation}

Now we are ready to estimate the average of the function (\ref{212}).  Using (\ref{estimin}) and the positiveness of the
 function (\ref{212}) for real $\lambda$ and $\varPhi(\lambda)$ for
$\lambda\in(\lambda_0-2\varepsilon,\lambda_0+2\varepsilon)$, we get
$$
\frac{1}{k_n}\int_{\lambda_0-\varepsilon}^{\lambda_0+\varepsilon}
\Im u_\infty(p_0)d\lambda \leq
\frac{1}{k_n}\int_{\lambda_0-\varepsilon}^{\lambda_0+\varepsilon}
\varPhi(\lambda)\Im u_\infty(p_0)d\lambda
\leq
\frac{1}{k_n}\int_{\lambda_0-2\varepsilon}^{\lambda_0+2\varepsilon}
\varPhi(\lambda)\Im u_\infty(p_0)d\lambda
$$
\begin{equation}\label{ineqRL}
=\frac{1}{k_n}\int_{\lambda_0-2\varepsilon}^{\lambda_0+2\varepsilon}
\varPhi(\lambda)\Im u^0_\infty(p_0)d\lambda-
\frac{1}{k_n}\int_{\lambda_0-2\varepsilon}^{\lambda_0+2\varepsilon}
\varPhi(\lambda)\Im f_\lambda(p_0)d\lambda.
\end{equation}

From (\ref{auxRI1}), the integrand of the first integral in the right hand side of (\ref{ineqRL}) does not
exceed $C\varepsilon$, i.e., the first term does not
exceed $C\varepsilon^2$, where the constant $C$ is independent of
$\varepsilon$ and $k$.

Let us study the second part in (\ref{ineqRL}). Since
$\varPhi(\lambda)$ is a real-valued function for real $\lambda$, we have
\begin{equation}\label{lll3}
\int_{\lambda_0-2\varepsilon}^{\lambda_0+2\varepsilon} \varPhi(\lambda)\Im
f_\lambda(\theta_0) d\lambda= \Im
\int_{\lambda_0-2\varepsilon}^{\lambda_0+2\varepsilon}
\varPhi(\lambda)f_\lambda(\theta_0) d\lambda.
\end{equation}

\begin{figure}[h]
\begin{picture}(1.5,70)
\scalebox{0.9}{ \rput(8,0.2){

\psdot[dotstyle=*](0,0)

 \rput(0,-0.4){\small$\lambda_0$}
 \rput(2,-0.4){\small$\lambda_0+2\varepsilon$}
 \rput(-2,-0.4){\small$\lambda_0-2\varepsilon$}

 \rput(-4.3,2.4){$\Im \lambda$}
 \rput(-4.3,2.0){$2 \varepsilon$}

 \rput(0,2.4){\small$\Gamma_2$}
 \rput(2.4,1){\small$\Gamma_3$}
 \rput(-2.4,1){\small$\Gamma_1$}

\psline{->}(-5,0)(5,0)
\pspolygon[linewidth=1pt,showpoints=true](-2,0)(-2,2)(2,2)(2,0)
\psline[linecolor=gray,linestyle=dashed](-4,2)(-2,2)

\psline{->}(-4,-0.5)(-4,2.5)

 \rput(4.7,-0.5){$\Re \lambda$}
} }
\end{picture}
 \caption{Contour $\Gamma$}
\label{fig6}
\end{figure}
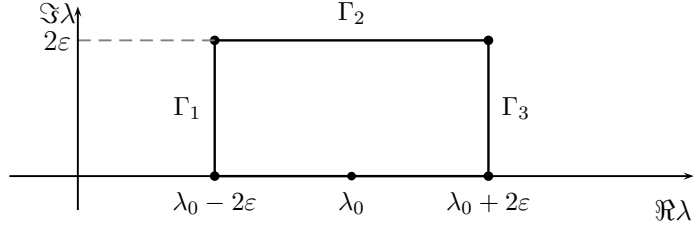

One can easily see that the Kirchhoff approximation $u^0$ defined
by (\ref{appEik2}) is analytic in $\lambda$. Then from
(\ref{FarFieldExpres}) it follows that $u^0_{\infty}$ is analytic
in $\lambda$. The function  $u_{\infty}$ is also analytic in
$\lambda,~\Im\lambda\geq 0,$ (see eg \cite{dtn}, \cite{vainLak}).
Thus, $f_\lambda(\theta_0)$ is analytic when $\Im\lambda\geq 0$,
and the contour of integration in (\ref{lll3}) can be replaced by
the contour $\Gamma= \Gamma_1 \cup \Gamma_2 \cup \Gamma_3$ defined in
Fig.~\ref{fig6}.

From here and (\ref{ineqRI}) it follows that
\[
\left | \int_{\lambda_0-2\varepsilon}^{\lambda_0+2\varepsilon}
\varPhi(\lambda)f_\lambda(\theta_0) d\lambda \right | = \left |
\int_{\Gamma}\varPhi(\lambda) f_\lambda(\theta_0)d\lambda \right |
\leq o(k)\max_{\Gamma}
 \frac{|\varPhi(\lambda) | }{\Im\lambda}  |\Gamma  |.
\]
One can show that $\max_{\Gamma} \frac{|\varPhi(\lambda)|}{\Im
\lambda} \leq \frac{C}{\varepsilon}.$
 Since $|\Gamma|=8\varepsilon$, we obtain
the following relation which holds for the second term in the
right-hand side of (\ref{ineqRL}). There exists $\varepsilon_0>0$
such that
$$
\delta_n:=\max_{0<\varepsilon<\varepsilon_0}\frac{1}{k_n}\left |
\int_{\lambda_0-2\varepsilon}^{\lambda_0+2\varepsilon}\varPhi(\lambda)
f_\lambda(\theta_0)d\lambda \right |\rightarrow 0,
\quad k_n \rightarrow \infty.
$$
 Now from (\ref{ineqRL}) and (\ref{212}) we get that, for  $\varepsilon \in (0,\varepsilon_0)$,
$$
\frac{1}{2\varepsilon}\int_{\lambda_0-\varepsilon}^{\lambda_0+\varepsilon}
\sigma_\lambda(k_n) d\lambda \leq C\varepsilon +
\frac{\delta_n}{\varepsilon}, \quad k_n \rightarrow \infty.
$$

The choice $\varepsilon_n=\sqrt{\delta_n}$ implies the statement of Theorem \ref{tint1}. The Consequence from the theorem
 follows if $\lambda_0$ is chosen to be the minimum value of $\sigma_\lambda(k_n)$ on the interval $\lambda_0-\varepsilon_n\leq \lambda \leq \lambda_0+\varepsilon_n$.

\section{Appendix I}

Here we prove Lemma \ref{EIKlemma0e}. The proof is a direct consequence of formula (\ref{impedanceCompostionDef})
 and lemmas \ref{lemma1} and \ref{lemma2} below concerning the asymptotic behavior of the simple and double layers as $k\rightarrow\infty, ~r$ is bounded.
These asymptotics are different in the shadow zone, the reflected zone and in the complementary region.

We show that on $M$ and in shadow zone we have
\begin{equation}\label{simM1}
D(r)=\frac{2\pi}{ikn_{\alpha}}e^{ik(\alpha,r)}+O(k^{-3/2}),~~r \in M\cup S(\alpha),~~k\rightarrow\infty.
\end{equation}
If $r$ belongs to the reflected zone, then
\begin{equation}\label{asref}
D(r)=\frac{2\pi}{ik n_\alpha} e^{ikt_0} e^{ik(\alpha^*\cdot
r)}+ O(k^{-3/2}) , \quad r
\in R(\alpha),~~k\rightarrow\infty,
\end{equation}
where constant $t_0=t_0(\alpha,M)$ defined from the continuity on $M$, i.e.,
\begin{equation}\label{CTransmission}
e^{ikt_0}e^{ik(\alpha^*\cdot r)}=e^{ik(\alpha \cdot r)}, \quad r
\in M.
\end{equation}
If $r$ is in the complement to those zones, then
\begin{equation}\label{simMb1}
D(r)=O(k^{-3/2}),~~k\rightarrow\infty,
\end{equation}
However, these asymptotics are not uniform, and exact statement is as follows.
\begin{lemma}\label{lemma1}
1. Expansion (\ref{simM1}) is valid uniformly on any compact subset of $M$ and on any compact set strictly inside of the
 shadow zone. Expansion (\ref{asref}) is valid uniformly on any compact set strictly inside of the reflected zone.

2. Estimate (\ref{simMb1}) is valid uniformly on any compact set which does not have common points with the closure of the
 shadow or illuminated zones.

3. The above expansions on compact sets strictly inside $S(\alpha)$ or $ R(\alpha)$, or strictly outside of these zones
can be differentiated any number of times with the remainders multiplied by $k$ after each differentiation.

4. Derivatives of $D(r)$ of any order in directions tangential to $M$ are continuous on $M$, and the expansion (\ref{simM1})
on $M$ can be differentiated  any number of times in those directions (with the corresponding increase of the order of the
 remainders). The normal derivative of $D(r)$ has a jump on $M$, and
\begin{equation}\label{nder}
\frac{\partial}{\partial n}D(r)=-2\pi\eta e^{ik(\alpha,r)},~~ r \in M,
\end{equation}
where the left hand side is defined according to (\ref{dnD})

5. The following uniform bound is valid on any bounded region of $R^3$
\begin{equation}\label{simMb2}
|D(r)|\leq \frac{C(R_0)}{k},~~|r|<R_0,~k \rightarrow\infty ,
\end{equation}
\begin{equation}\label{simMb3}
|P_mD(r)|\leq C(R_0)k^{m-1},~~r\notin \overline{M},~|r|<R_0,~k \rightarrow\infty,
\end{equation}
where $P_m$ is an arbitrary partial derivative of order $m$.
\end{lemma}

\textbf{Proof.} First, we discuss the smoothness of $D(r)$ on $M$ (obviously $D\in C^{\infty}$ when $r\notin \overline{M}$). Since  the integrand in (\ref{DDef}) has a weak singularity, $D(r)$ is continuous on $\overline{M}$. Further, let $r\notin \overline{M},~q\in M$. For any unit vector $v$ orthogonal to $n$ (i.e. $v$ is in the plane containing $M$), we have
$\frac{\partial}{\partial v} |r-q|=-\frac{\partial}{\partial v_q}|r-q|$ where
 $\frac{\partial}{\partial v},~\frac{\partial}{\partial v_q}$ are derivatives in the direction of
 the vector $v$ with respect to $r$ and $q$, respectively. Thus, after integration by parts we get
\[
\frac{\partial}{\partial v}D(r)=\int_M\eta e^{ik(\alpha,q)}\frac{\frac{\partial}{\partial v}e^{ik|r-q|}}{|r-q|}dS_{q}-\int_M\eta e^{ik(\alpha,q)}e^{ik|r-q|}\frac{\partial}{\partial v_q}[\frac{1}{|r-q|}]dS_{q}
\]
\begin{equation}\label{tander}
=\int_M \frac{\eta}{|r-q|}(\frac{\partial}{\partial v}+\frac{\partial}{\partial v_q}) e^{ik((\alpha,q)+|r-q|)}dS_{q}+H^{(v)}=
ik(\alpha,v)D(r)+H^{(v)},
\end{equation}
where $H^{(v)}=\int_{ M}(\frac{\partial}{\partial v_q}\eta) \frac{e^{ik((\alpha,q)+|r-q|)}}{|r-q|}dS_{q}.$
Obviously $H^{(v)} $ is continuous, and so is $\frac{\partial}{\partial v}D(r)$. If the vector $u$ is also orthogonal to $n$, then the same arguments imply
\[
\frac{\partial^2}{\partial u\partial v}D(r)=-k^2(\alpha,u)(\alpha,v)D(r)+ik(\alpha,v)H^{u}+ik(\alpha,u)H^{v}
\]
\begin{equation}\label{tander1}
+\int_{ M}\frac{\partial^2\eta}{\partial u_q\partial v_q}
\frac{e^{ik((\alpha,q)+|r-q|)}}{|r-q|}dS_{q}.
\end{equation}
This provides continuity of second derivatives. The higher order derivatives can be treated similarly.

The normal derivative of a single layer has a jump on the surface of integration. This fact and the limiting values of the
 normal derivatives can be found in the many textbooks on partial differential equations. In our case the limit (\ref{nder}) of the normal
derivative is proportional to the density and does not contain an integral term since the surface of integration is flat.
Formula (\ref{nder}) can also be treated as a simple exercise in the distribution theory.

 Our next goal is to find the asymptotics of $D(r),~k\rightarrow\infty,$ on a compact subset $K\subset M$. Note that the
 integrand in (\ref{DDef}) is not smooth in this case, and the stationary phase method cannot be applied.
Let us introduce polar coordinates $(\rho, \phi)$ on the plane $P$ with the center at the point $r$ and the polar angle
 being zero along of the ray $\tau$ whose direction coincides with the direction of the projection $\widehat{\alpha}$ of
the vector $\alpha$ on $P$. If $\alpha$ is orthogonal to $P$ (i.e. $\widehat{\alpha}=0$), then $\tau$ can be chosen
arbitrarily. Polygon $\partial M$ in polar coordinates has the form $\rho = \rho (\phi,r), ~0 \leq\phi <2\pi$, where $\rho$
 as a function of $\phi$ is continuous, $2\pi$-periodic and analytic for all the values of $\phi$ except those which
correspond to the vertices of $\partial M$.
We write the integral $D(r)$ in polar coordinates and integrate by parts in $\rho$. When $r\in K\subset M$ and $k$ is
large enough so that $\eta=1$ on $K$, we get
\[
D(r)=e^{ik(\alpha,r)}\int_{0}^{2\pi}\int_{0}^{\rho(\phi,r)}\eta e^{ik\rho (1+|\widehat{\alpha}|\cos\phi )}d\rho d\phi,
\]
\begin{equation}\label{ss}
=\frac{e^{ik(\alpha,r)}}{ik}(\int_{0}^{2\pi} \frac{-1}{1+|\widehat{\alpha}|\cos\phi } d\phi-\int_{0}^{2\pi}\int_{0}^{\rho(\phi,r)}\eta_{\rho} \frac{e^{ik\rho(\phi,r) (1+|\widehat{\alpha}|\cos\phi )}}{1+|\widehat{\alpha}|\cos\phi }d\rho d\phi),
\end{equation}
Since vector $\alpha$ is transversal to $P$, $|\widehat{\alpha}|<1$, and
$\int_0^{2\pi}\frac{d\phi}{1+|\widehat{\alpha}|\cos\phi}=\frac{2\pi}{\sqrt{1-|\widehat{\alpha}|^2}}
=\frac{-2\pi}{n_{\alpha}},$
formula (\ref{ss}) implies
\begin{equation}\label{DonM}
D(r)=\frac{1}{ik}e^{ik(\alpha,r)}(\frac{2\pi}{n_{\alpha}}-\int_{0}^{2\pi}\int_{0}^{\rho(\phi,r)}\eta_{\rho}\frac{e^{ik\rho(\phi,r) (1+|\widehat{\alpha}|\cos\phi )}}{1+|\widehat{\alpha}|\cos\phi }d\rho d\phi ),~~r\in K\subset M.
\end{equation}
The interior integral, above after integration by parts $m$ times, takes the form
\[
(\frac{-1}{ik})^m\int_{0}^{\rho(\phi,r)}\frac{\partial^{m+1}\eta}{(\partial\rho)^{m+1}}\frac{e^{ik\rho(\phi,r) (1+|\widehat{\alpha}|\cos\phi )}}{(1+|\widehat{\alpha}|\cos\phi )^{m+1}}d\rho=O(k^{(1-3m)/4}),~~k \rightarrow\infty.
\]
The latter relation follows from (\ref{eta}). Hence, for $r\in K\subset M$, (\ref{DonM}) implies (\ref{simM1}).

The tangential derivatives of $D(r)$ can be evaluated using (\ref{tander}), (\ref{tander1}) and similar formulas for the
derivatives of higher order. One only needs to note that functions similar to $H^{(v)}$ have order $O(k^{-\infty}),
~k\rightarrow\infty$, i.e., they decay at infinity faster than any power of $k$.
This can be justified precisely as the similar statement above for the integral in (\ref{DonM}).
Thus, the asymptotics (\ref{simM1}) on $M$ can be differentiated in tangential directions.

We now consider the asymptotics of $D(r)$ and its derivatives on compact sets outside of the boundaries of the shadow and reflected zones. If $r\notin\overline{M}$, the integrand in (\ref{DDef}) is smooth, and the stationary phase method can be applied. Let $F=F(q)$ be the phase function in (\ref{DDef}). Since $\nabla_q F=\alpha-\frac{r-q}{|r-q|},$
the stationary phase points are defined by the equation
\begin{equation}\label{stp}
\alpha-\frac{r-q}{|r-q|}=cn,\text{where}~~c \text{ is a constant},~~q \in M.
\end{equation}
Both vectors on the left are unit vectors, and therefore, (\ref{stp}) holds if and only if
$
r-q=c_1\alpha$ or $r-q=c_2\alpha^*$, where $c_i=|r-q|\geq 0, ~~q \in M.
$
This is possible only if $r$ is in the shadow or reflected zone, respectively. The stationary point is on $\partial M$ if
and only if $r$ belongs to the lateral boundary of these zones. If $r$ belongs to a compact set whose intersection with the
boundaries of those zones is empty, the stationary phase point is strictly inside $M$ or outside of $\overline{M}$. Thus,
the stationary phase method implies the uniform validity of the expansions  (\ref{simM1}), (\ref{asref}), (\ref{simMb1})
for $D$ and its derivatives on these compact sets.

Since the amplitude factor $\eta(k,q)/|r-q|$ in the integrand (\ref{DDef}) depends on $k$, we will justify the latter
statement a little more rigorously. Consider a compact set $J$ strictly inside of the reflected zone. Let $J'$ be the
projection of $J$ into $M$ parallel to the vector $\alpha^*$. Then $J'$ is located strictly inside $M$, and the stationary
phase point belongs to $J'$ when $r\in M$. Let $\varsigma=\varsigma(r)\in C^{\infty}(M)$ be an infinitely smooth function
 which is equal to one on $J'$ and
zero in some neighborhood of $\partial M$. Let $k$ be so large that $\eta=0$ on the support of $\varsigma$. We write $D(r)$
 as the sum $D_1(r)+D_2(r)$ where $D_1,D_2$ are given by (\ref{DDef}) with additional factors $\eta,1-\eta$ in the
integrand, respectively. The stationary phase method can be applied to $D_1$, since the amplitude factor for that integral
does not depend on $k$. In order to complete the proof of (\ref{asref}) for $r\in J$ it remains to show that
$D_2=O(k^{-\infty}),~k\rightarrow \infty$. If $f(k,q)=\eta(1-\varsigma)/|r-q|$, then
\[
D_2=\int_Mfe^{ikF}dS_q=\int_M\frac{f}{|\nabla F|^2}(\nabla F\cdot\nabla F)e^{ikF}dS_q=-\frac{1}{ik}\int_M\nabla(\frac{f\nabla F}{|\nabla F|^2})e^{ikF}dS_q.
\]

This integration by parts can be repeated as many times as we please. This combined with (\ref{eta}) provides the estimate
 for $D_2.$ Thus, (\ref{asref}) for $r\in J$ is proved. Estimates for the compact sets in the shadow zone and outside of the shadow and reflected
zones are treated similarly.

It remains to prove the last statement of Lemma \ref{lemma1}. We begin with the proof of (\ref{simMb2}). The proof
will be based on the same arguments which were used above in order to justify the asymptotics of $D(r)$ on $M$. Let us
assume first that $r$ belongs to the reflected zone, i.e. $r=q_0+t\alpha^*,~q_0 \in M, t>0$. We introduce polar
 coordinates $(\rho,\phi)$ on the plane $P$ with the origin at $q_0$ and the polar angle defined in the third paragraph of the proof (where polar coordinates were used). Then similar to
the first equality (\ref{ss}), we have
\begin{equation}\label{polc}
D(r)=e^{ik(\alpha,q_0)}\int_{0}^{2\pi}\int_{0}^{\rho(\phi,q_0)}\frac {\eta e^{ikf(\rho,t,\phi)}}{ g(\rho,t,\phi)}\rho d\rho d\phi,
\end{equation}
where
$
f=\rho |\widehat{\alpha}|\cos\phi +\sqrt{t^2- 2t|\widehat{\alpha}|\rho\cos\phi +\rho^2},~~g= \sqrt{t^2- 2t|\widehat{\alpha}|\rho\cos\phi +\rho^2}.
$

For any function $u(\rho,t,\phi),$ denote by $u'$ its derivative with respect to $\rho$. Let $h(\rho,t,\phi)=f'(\rho,t,\phi)/\rho$. Since $|\widehat{\alpha}|<1$, one can easily check that $h(\rho,1,\phi)\in C^{\infty},~h(\rho,1,\phi)\neq 0$ and $0<c_1\rho^{-1}<h<c_2\rho^{-1},~ \rho \rightarrow\infty$. Moreover,
 $
   1/|hg|,~|(hg)'|\leq C<\infty$ when $t=1,~\rho>0.
  $
  Then from the homogeneity of $f$ and $g$ with respect to $(\rho,t)$ it follows that the same estimates
(with the same constant $C$ ) hold for all $t>0$:
\begin{equation}\label{321}
   \frac{1}{|hg|},~|(hg)'|\leq C<\infty,~~t,\rho>0.
\end{equation}

Using integration by parts, we get:
\begin{eqnarray*}
D(r)=e^{ik(\alpha,q_0)}\int_{0}^{2\pi}\int_{0}^{\rho(\phi,q_0)}\frac {\eta}{ h(\rho,t,\phi)g(\rho,t,\phi)}f'(\rho,t,\phi)e^{ikf(\rho,t,\phi)} d\rho d\phi
\\
=(ik)^{-1}e^{ik(\alpha,q_0)}\int_{0}^{2\pi}[\frac {\eta}{ h(\rho,t,\phi)g(\rho,t,\phi)}e^{ikf(\rho,t,\phi)}]|_{\rho=0} d\phi
\end{eqnarray*}
\begin{equation}\label{uniest}
-(ik)^{-1}e^{ik(\alpha,q_0)}\int_{0}^{2\pi}\int_{0}^{\rho(\phi,q_0)}(\frac {\eta '}{ hg}+\eta(\frac{1}{hg})')e^{ikf(\rho,t,\phi)} d\rho d\phi.
\end{equation}
We now split the last term into $I_1+I_2$ with factors $\eta'$ and $\eta$ in the integrands, respectively. Formulas
(\ref{321}) justify (\ref{simMb2}) for the first term on the right hand side of (\ref{uniest}) and for $I_2$. They also lead to the estimate $I_1<Ck^{-1}\int_M|\eta '|dS_q$ which implies  (\ref{simMb2}) for $I_1$, since $|\eta '|=O(k^{1/4})$ and the support of $\eta '$ has measure of order $O(k^{-1/4})$. The proof of  (\ref{simMb2}) when $r$ is in the reflected zone is complete.

 The same arguments are valid if $r$ in the same half space bounded by $P$, but does not belong to the closure of the reflected zone. Then $q_0$ does not belong to $M$. The limits of integration in (\ref{polc}) will be different, namely, $\phi \in I, ~\rho_1(\phi,q_0 )<\rho <\rho_1(\phi,q_0 )$. Here $\rho_1, \rho_2$ are values of $\rho$ where the ray from the origin with the polar angle $\phi$ intersects $\partial M$, and $I$ is the set of values of the angle for which such an intersection is not empty. This leads to (\ref{uniest}) where the first term on the right has to be omitted and the limits of integration in the second term have to be changed. Obviously, estimate (\ref{simMb2}) is still valid. In order to obtain (\ref{simMb2}) when $r$ is in another half space, one needs only to replaced $\alpha^*$ by $\alpha$.

Let us establish the validity of (\ref{simMb3}). We begin with the case of $P_m=D_u$, i.e. with the estimate of a
tangential derivative of the first order. The desirable estimate follows from (\ref{tander}): the first term in the
right-hand side of (\ref{tander}) satisfies  (\ref{simMb3}) due to (\ref{simMb2}), and the second term has an even more
refined estimate. In fact, function $H^{(u)}$ has the same form as $D(r)$, but with $\eta$ replaced by its derivative. Thus the
arguments justifying (\ref{simMb3}) for $D(r)$ will provide the same estimate for $H^{(u)}$ with an extra factor $k^{1/4}$ in the right-hand side due to (\ref{eta}). Tangential derivatives of $D(r)$ of higher order can be obtained similarly.

We now estimate $\frac{\partial}{\partial n}D(r)$. Let $r_0$ be the orthogonal projection of $r$ onto the plane $P$, i.e. $r=r_0+nt,~r_0\in P,~ 0\neq t\in \mathbb R.$ Then $\frac{\partial}{\partial n}D(r)=I_1+ikI_2,$
where
\[
I_1=\int_M\eta\frac{-t}{|r-q|^3}e^{ik((\alpha\cdot q)+|r-q|)}dS_q+\int_M\eta\frac{t}{|r-q|^2}e^{ik((\alpha \cdot q)+|r-q|)}dS_q.
\]
The second factor differs from $D(r)$ only by an extra factor $\frac{t}{|r-q|}$ in the integrand. This factor is bounded and homogeneous  in $(r,q)$ of zero order. Thus, the proof of (\ref{simMb2}) for $D$ (see (\ref{polc})-(\ref{uniest})) can be repeated for $I_2$, i.e. $|I_2|<C/k,~k\rightarrow\infty$. $I_1$ can be estimated very easily:
if $r_0^2+t^2<R_0^2,$ ($|r|$ is bounded), then by using polar coordinates centered at $r_0$ we obtain
\[
|I_1|\leq C\int_{|q-r_0|<c}\frac{|t|}{((r_0-q)^2+t^2)^{3/2}}dS_q
\leq 2\pi C\int_{0}^c\sigma(\frac{|t|}{(\sigma^2+t^2)^{3/2}}\leq C_1,
\]
i.e. (\ref{simMb3}) holds for $\frac{\partial}{\partial n}D(r)$.

If $P_m$ contains differentiation of order $m-1$ in tangential directions and the derivative of the first order in the
 normal to $M$ direction, then (\ref{simMb3}) can be justified by a combination of the arguments used to prove (\ref{simMb3})
 for tangential derivatives and for $\frac{\partial}{\partial n}D(r)$: one needs to start with (\ref{tander}) or similar
 formula  for the tangential derivatives of higher order, then apply $\partial/\partial n$ and repeat the arguments used to
 estimate $\frac{\partial}{\partial n}D$. Finally, note that
$(\triangle+k^2)D(r)=0$ when $r\notin \overline{M}$. Thus
\begin{equation}\label{DN}
\frac{\partial^2}{\partial n^2} D(r) = \left (\Delta_M + k^2 \right
) D(r),~~r \notin M,
\end{equation}
where $\Delta_M$ is the two-dimensional Laplacian on $M$. Hence $P_mD(r)$ for arbitrary $P_m$ can be expressed through
derivatives of $D$ containing differentiation in the normal direction of at most first order. This proves  (\ref{simMb3})
for arbitrary $P_m$.

  The proof of Lemma \ref{lemma1} is complete.

The double layer potential $N(r)=N_{\alpha, M}(r)$ is given by (\ref{NDef}), (\ref{dnN}). Obviously, since  $M$ is
flat, $N(r)=-\frac{\partial}{\partial n}D(r)$. Thus, the next lemma about the properties of $N(r)$ is a direct consequence of Lemma \ref{lemma1} and (\ref{DN}).

\begin{lemma}\label{lemma2}
1. If $r \in  M$ and $k \rightarrow \infty$, then
$$
N(r)={2\pi}  e^{ik(\alpha \cdot r)},~~
\frac{\partial}{\partial n} N(r) =-{2\pi i k}(n \cdot
\alpha) e^{ik(\alpha \cdot r)}(1+ O(1/\sqrt k) ),
$$
where the estimate of the remainder is uniform on any compact subset of $ M$ and the remainder is bounded on $ \overline{M}$, i.e,
$\left |\frac{\partial}{\partial n} N(r) \right | \leq
Ck, \quad r \in M,~~k \rightarrow\infty.$

2. Let $K_1, K_2, K_3$ be compact sets strictly inside of the shadow zone or reflected zone, or strictly outside of
 these zones, respectively, and let $t_0$ be defined by (\ref{CTransmission}). Then the following expansions are uniformly valid  on these compact sets.
 \begin{eqnarray}
N(r)=-{2\pi} e^{ik(\alpha \cdot r)}+O(1/\sqrt k),~~r\in K_1\subset S(\alpha) ,~~ k \rightarrow \infty,
\\
N(r)={2\pi}e^{ikt_0} e^{ik(\alpha^* \cdot r)}+
O(1/\sqrt k), ~~r\in K_2\subset R(\alpha) ,~~ k \rightarrow
\infty,
\\
N(r)=O(1/\sqrt k) ,~~r \in K_3 \subset\mathbb R^3 \backslash \overline{(\I(\alpha) \cup
\LL(\alpha))}, \quad k \rightarrow
\infty
\end{eqnarray}
These expansions on $K_i$ can be differentiated any number of times with the remainders multiplied by $k$ after each differentiation.

3. A uniform bound of $N(r)$ is valid in any bounded region:
\begin{equation}\label{uniformNS}
|N(r)| <C(R_0),~~ |\nabla N(r)| <kC(R_0), ~~|r|<R_0,~ r\notin \overline{M},\quad k \rightarrow \infty.
\end{equation}

\end{lemma}

\end{document}